\documentclass[fleqn,usenatbib]{mnras}

\usepackage{newtxtext} 

\usepackage[T1]{fontenc}

\DeclareRobustCommand{\VAN}[3]{#2}
\let\VANthebibliography\thebibliography
\def\thebibliography{\DeclareRobustCommand{\VAN}[3]{##3}\VANthebibliography}

\usepackage{graphicx}	
\usepackage{amsmath}	
\usepackage{amssymb}	

\newcommand{\Msun}{M$_{\odot}$}

\newcommand{\Gr}{\mathrm{G}}

\title[Gas fraction, feedback and giant clump evolution]{The role of gas fraction and feedback in the stability and evolution of galactic discs: implications for cosmological galaxy formation models}

\author[Fensch J. \& Bournaud F.]{
J\'er\'emy Fensch$^{1,}$\thanks{email: jeremy.fensch@ens-lyon.fr}
and Fr\'ed\'eric Bournaud$^{2}$
\\
$^{1}$Univ. Lyon, ENS de Lyon, Univ. Lyon 1, CNRS, Centre de Recherche Astrophysique de Lyon, UMR5574, 69007 Lyon, France\\
$^{2}$AIM, CEA, CNRS, Universit\'e Paris-Saclay, Universit\'e Paris Diderot, Sorbonne Paris Cit\'e, 91191 Gif-sur-Yvette, France
}

\date{Accepted XXX. Received YYY; in original form ZZZ}

\pubyear{2021}

\begin{document}
\label{firstpage}
\pagerange{\pageref{firstpage}--\pageref{lastpage}}
\maketitle

\begin{abstract}
High-redshift star-forming galaxies often have irregular morphologies with {\it giant clumps} containing up to $10^{8-9}$ solar masses of gas and stars. The origin and evolution of giant clumps are debated both theoretically and observationally. In most cosmological simulations, high-redshift galaxies have regular spiral structures or short-lived clumps, in contradiction with many idealised high-redshift disc models. Here we test whether this discrepancy can be explained by the low gas fractions of galaxies in cosmological simulations. We present a series of simulations with varying gas fractions, from 25\%, typical of galaxies in most cosmological simulations, to 50\%, typical of observed galaxies at 1.5 < z < 3. We find that gas-poor models have short-lived clumps, that are unbound and mostly destroyed by galactic shear, even with weak stellar feedback. In contrast, gas-rich models form long-lived clumps even with boosted stellar feedback. This shows that the gas mass fraction is the primary physical parameter driving violent disc instabilities and the evolution of giant clumps on $\sim$10$^8$~yr timescales, with lower impact from the calibration of the stellar feedback. Many cosmological simulations of galaxy formation have relatively gas-poor galactic discs, which could explain why giant clumps are absent or short-lived in such models.
Similar baryonic and dark matter mass distribution could produce clumpy galaxies with long-lived clumps at $z\sim2$ if the gas fraction was in better agreement with observations.
\end{abstract}


\section{Introduction}

Star-forming galaxies at the peak of the cosmic star formation history \citep[redshift $1.5 < z < 3$, e.g ][]{Madau14}, are observed to have a typical morphology different from that of their local counterparts. Their rest-frame optical and near-UV morphology is often not dominated by spiral arms, but rather irregular with a few large ($\sim$\,0.5\,kpc) and massive ($\sim$\, $10^{8-9}$\,M$_\odot$) stellar over-densities, called {\it giant clumps} \citep[see e.g.][]{Cowie95, Elmegreen07, Guo18, Zanella19}. In the most clumpy galaxies, up to 50\% of the SFR and 10\% of the stellar mass can be contained by the clumps \citep{Wuyts12, Soto17, Guo18}. Observed stellar population in giant clumps have ages around 100-200~Myr \citep{Wuyts12}. It has been proposed that clumps can migrate inward through dynamical friction \citep{Noguchi99, Elmegreen05} if they are long-lived against stellar feedback. Radial age gradients that are observed among giant clumps are consistent with inward migration of long-lived clumps \citep{Forster11, Guo12, Guo15}. Giant clumps are usually proposed to form out of violent disc instabilities due to the high gas content of high-redshift galaxies \citep[see e.g.][]{Elmegreen08, Dekel09, Agertz09, Inoue16}, with a smaller fraction being accreted through minor mergers \citep{Mandelker14, Zanella19}.

One should note that the nature and properties of clumps from observational data is debated. For instance, observations of strongly lensed galaxies \citep[see e.g.][]{Cava18, DEssauges-Zavadsky19} detected only small and low mass clumps. Furthermore, observations with the Atacama Large
Millimeter Array (ALMA) did not find counterparts of the optical detections in the gas component \citep{Rujopakarn19, Ivison20}. This could be interpreted in terms of efficient gas consumption or expulsion, but there could also be observational bias against detecting large structures in this type of observations: in particular, giant clumps likely contain numerous smaller sub-clumps that are resolved and detected in lensed observations and/or high-resolution ALMA datasets \citep{Behrendt16, Faure21}. 

On the theoretical side, formation and evolution of the clumps are also subject of an active debate. On the one hand, several numerical simulations suggest that the clumps are disrupted by stellar feedback in short timescales, that is at most a few tens of Myr \citep{Genel12, Tamburello15, Oklopcic17}, or have no physical meaning at all \citep[see e.g.][]{Buck17}. Massive star-forming galaxies at $z\sim 2$ in the Illustris-TNG simulations rarely or never exhibit giant clumps, but are almost always dominated by a regular spiral structure, instead \citep[e.g., ][ Fig. 6 and 7]{Pillepich19}.

 On the other hand, several simulations including a thorough model of stellar feedback (supernovae, radiation pressure, ionization by HII regions) do show the formation of relatively long-lived (> 100-500 Myr) clumps \citep{Perret14, Bournaud14, Ceverino14, Fensch17}. For instance, \citet{Perez13} show that clumps can survive even when strong outflows -- consistent with observations \citep{ Genzel11, Newman12} -- are launched from their star formation activity. Yet, simulations predicting long-lived clumps are mostly idealized galaxy-scale simulations, while the majority of cosmological simulations predict either short-lived clumps or low-mass clumps \citep[but see][]{Agertz09, Mandelker14, Dubois20}. The two factors discriminating between these models seem to be stellar feedback modelling and gas mass fraction. While these two parameters are likely degenerate across the whole history of a galaxy, stellar feedback impacting the rhythm at which gas is consumed and expelled, on a $10^{8}$~Myr timescale stellar feedback and gas mass fraction might have a different impact on the stability and evolution of simulated disks. One should note that idealised and cosmological simulations do not calibrate their feedback recipes in the same way. While the former can be calibrated using resolved quantities, such as the outflow rates or molecular clouds lifetimes, in most of the cosmological simulations feedback is calibrated to reproduce the stellar-to-halo mass at z=0 \citep[see e.g.][]{Hopkins14, Keller15, Sokolowska17}. \\
 
A recurring issue in state-of-the-art cosmological simulations could be that the gas fraction in massive star-forming galaxies at $z$=1-3 is relatively low. Star-forming galaxies at $z\simeq 2$ with stellar masses of a few $10^{10}$ to $10^{11}$\,M$_{\sun}$ in the FIRE simulation sample typically have gas fraction\footnote{the gas fraction being the fraction of the baryonic mass in the form of gas, regardless of the dark matter mass.} in the 10--30\% range, with an average of the order of 20\% \citep[see their Fig.~12]{ Feldmann17}. Similar gas fractions are measured for the same type of galaxies in the Illustris \citep{Genel14} and Illustris-TNG \citep[ see their Fig.~17]{Pillepich19} simulations, as well as in other simulations series \citep[see e.g.][and references therein]{Lagos15, Popping19}. The details analysis of a massive $z\sim2$ galaxy with short-lived clumps in a cosmological simulation presented in \citet[][hereafter O17]{Oklopcic17} has a gas mass fraction around 25\% (see their Fig.~5). In contrast, observations of star-forming galaxies in the same mass and redshift range concur to a gas fraction of 50\%, based on a large variety of gas tracers (CO: \citealt{Daddi10,Tacconi10,Tacconi18} -- CI: \citealt{Valentino20} -- CII: \citealt{Zanella18} -- dust: \citealt{Magdis12, Santini14}). Observed samples are now large enough to constrain the average value of $\simeq$50\% by much less than a factor two. In addition, the typical value of $\simeq$25\% in cosmological simulation generally includes warm gas in the vicinity of the galaxy disk, while observations refer only to cold gas in the disk, generally even neglecting the likely presence of some warm gas: this can only strengthen the tension on the cold gas content of star-forming galaxies at $z \sim 2$ between cosmological simulations and observations. This discrepancy is also present in cosmological theoretical models \citep[see e.g.,][]{Dekel14}.

In this paper, we test whether the absence of long-lived giant clumps in many cosmological simulations could result from too low gas fraction rather than stellar feedback, and whether more realistic gas mass fractions would enable the formation of long-lived giant clumps for any plausible stellar feedback. Heuristically, at a given total baryonic mass and for a given rotation curve, a higher mass fraction in the dissipative gas component should make the disk more unstable (by lowering the total \citet{Toomre64} parameter) and promote the formation of massive clumps \citep{Noguchi98, Jog84, Rafikov01, Elmegreen05, Dekel09, Romeo10}.

The present article aims at quantifying the effect of the gas mass fraction on disc stability, taking as example a galaxy similar to that studied in the FIRE cosmological simulations in O17. The protocol consists in comparing the disc evolution for two galaxy models for which only the gas mass fraction is modified, i.e some stars from the disc is replaced by gas, to obtain two models with respective gas mass fraction of 25\% and 50\%. In particular, the total baryonic, dark matter and bulge masses and sizes are conserved. This enables to study the impact of this sole parameter while keeping other important parameters for disc instabilities, such as the shear and the local surface density. \\ 

The paper is structured as follows. In Section \ref{sec::sim}, we describe the numerical methods and simulation set. In Section \ref{sec::res}, we describe the formation of structures in the simulated discs. In Section \ref{sec::disc} and \ref{sec::conc} we respectively discuss the implication of our results and conclude.


\section{Simulations}
\label{sec::sim}

\subsection{Numerical methods}

We use the adaptive mesh refinement code RAMSES \citep{Teyssier02}. To ease the comparison with cosmological simulation setups, we use a refinement strategy similar to O17, which uses a smooth particle hydrodynamics method: a cell is refined if it contains more than 50 particles from initial conditions (dark matter and stars, named {\it old stars} thereafter), or if its mass of gas, dark matter, old stars and new stars is higher than $1.8 \times 10^{6} $~\Msun. We allow for a maximum spatial resolution of 20~pc and a maximum cell size of 320~pc. We set a background gas density equal to $2\times10^{-7}$ the density at the truncation radius of the disk, that is around $2\times10^{-7}$~cm$^{-3}$. 

The thermodynamical model includes heating and atomic cooling at solar metallicity and is fully similar to \citet{Renaud15}. We allow cooling down to 500~K.  Furthermore, we ensure that the Jeans length is resolved by at least four cells at the highest resolution, by introducing a numerical pressure through a temperature floor set by a polytrope equation of state at high density ($T \propto \rho^{2}$), which is called Jeans polytrope thereafter, which prevents numerical fragmentation \citep{Truelove97} by accounting for unresolved stabilizing turbulence support. We stress that, to allow the formation of clumps in galaxy simulations, one should have have spatial resolution elements smaller than a few times the Jeans length, i.e resolution smaller than 25-50 pc \citep[see e.g.][]{Teyssier10}, and have a cooling model which takes into account gas cooling below $10^{4}$~K, to allow gas to dissipate its kinetic energy \citep[see e.g.][]{Bournaud11}. These two conditions are met in our simulations.

Star formation happens for gas above a gas density threshold, $\rho_{0} = 10$~cm$^{-3}$, if its temperature is no more than $2\times10^{4}$~K above the Jeans polytrope temperature at the corresponding density. Gas is converted into stellar particles, called {\it new stars} thereafter, following a Schmidt law: $\dot{\rho}_{\star} = \epsilon_\mathrm{SF} (\rho_{\mathrm{gas}} / t_{\mathrm{ff}})$ \citep{Schmidt59, Kennicutt98}, with $\epsilon_\mathrm{SF}$ the efficiency per free fall time, and $ t_{\mathrm{ff}} = \sqrt{3\pi / (32G\rho)}$ the free-fall time.  $\epsilon_\mathrm{SF}$ is calibrated so that the isolated discs are located on the disc sequence of the Schmidt-Kennicutt diagram \citep{Daddi10a, Genzel10}, that is a star formation rate (SFR) of $ \simeq 40~$M$_{\odot}$/yr (resp. $ \simeq 60~$M$_{\odot}$/yr) in the low gas fraction case (resp. high gas fraction case) for each galaxy. One should note that the refinement criterion being based on the gas density, the effective resolution of the two simulations is different. Hence we use $\epsilon_\mathrm{SF}$=0.04 for the F25 models, and $\epsilon_\mathrm{SF}$=0.0025 for the F50 models. New stars have a mass proportional to $\epsilon_\mathrm{SF}$, which thus varies between simulations (250 and 4000 solar masses, respectively). This only changes the number of stellar particles formed in a given cell. We have checked that enforcing the same mass for new stellar particles, namely 4000 solar masses, does not significantly modify the evolution of the simulations, with less than 5\% variation in the 1~Myr-sampled cumulative stellar mass formed and a point-to-point standard deviation of 10\% in the 1~Myr-sampled star formation rate, similar to the intrinsic stochasticity of the simulation. Gas and stellar surface density maps of F50 simulations with two different masses for the stellar particles are shown in Appendix~\ref{App::stars}.

Three types of stellar feedback are modelled. The most realistic way to implement stellar feedback in galaxy simulation is still subject of debate in the community. Thus our sample of simulations uses three different calibrations of these types of feedback, detailed in Table~\ref{table::bffb}. These feedback calibrations are called {\it Weak}, {\it Medium} and {\it Strong}. This labelling is justified from the mass loading of galactic outflows generated by stellar feedback, as detailed in Section~\ref{sec::disc_fbk}. We stress that our feedback models are not initially calibrated to reproduce a given observable. Given the uncertainty on the effective energy deposition in the ISM at the 20~pc scale, we vary the remaining free parameters within reasonable boundaries to ensure that their values do not bias the results.

\begin{itemize}
    \item Photoionization from HII regions is modelled by heating up the gas in the \citet{Stromgren39} sphere to a temperature, $T_\mathrm{HII}$, which we set between $5 \times 10^{4}$~K and $2 \times 10^{5}$~K depending on the calibrations (See Table~\ref{table::bffb}). A detailed description of the numerical implementation is given in \citet{Renaud13}.
    \item Radiation pressure in HII regions is modelled via injection of a velocity kick for each cell in the \citet{Stromgren39} sphere. The photon  scattering factor is set between 2.5 and 7 depending on the calibration (See Table~\ref{table::bffb}). A detailed description of the numerical implementation is given in \citet{Renaud13}.
    \item Type-II supernovae (SN) are modelled via a combination of thermal and kinetic energy release. We assume that 20\% of the initial mass of our particles is in massive stars that will end their lives as SN, and will be released to the gas 10~Myr after the formation of the stellar particle. We inject a fraction of total energy E$_\mathrm{SN}$ = $10^{51}$~erg/10~M$_\odot$. The numerical implementation is described in \citet{Dubois08}. Depending on the calibration, between 80\% to 100\% of this energy is released thermally, and between 0\% and 20\% of this energy is released in kinetic form, the total energy released staying constant. It should be noted that thermal blasts are a relatively inefficient source of feedback at large scales because in dense gas this energy is quickly radiated away \citep[see e.g.][]{Martizzi15}. This is the reason why the {\it Medium} and {\it Strong} feedback calibrations presented in Table~\ref{table::bffb} trigger stronger outflows (see Sect.~4) than the {\it Weak} feedback calibration. The blast wave mass loading factor corresponds to the gas mass carried by the blast wave within the kinetic SN model, and is typically around 1 \citep[see][]{Dubois08}.
\end{itemize}

    \begin{table*}
    \centering
    \caption{Characteristics of the feedback used in the simulation set. These parameters are detailed in the text.}
    \label{table::bffb}
    \begin{tabular}{cccccc}
    \hline 
    Feedback & HII region  & Scattering  & SN thermal  & SN kinetic & Blast wave  \\ 
     calibration            & temperature    & factor   &  energy fraction  &  energy fraction &  mass loading\\ \hline 
    Weak             & $5 \times 10^{4}~\mathrm{K} $& 2.5                          & 100 \%       & 0\%                      & --                                 \\ \hline
    Medium          & $10^{5}~\mathrm{K}         $& 4                             & 92\%            &        8\%                   & 0.42                                  \\ \hline
    Strong           & $2\times 10^{5}~\mathrm{K}$  & 7                            & 80\%           & 20\%                     & 0.80                                 \\ \hline
    \end{tabular}
    \end{table*}

    \begin{table}
    \centering
    \caption{Characteristics of the galaxies used in the simulations. The relative gas and stellar masses are chosen to have a baryonic gas mass fraction of 25\% and 50\% for the models F25 and F50 respectively. The relatively low mass of the dark matter halo comes from the truncation procedure.}
    \label{table::galaxies}
    \begin{tabular}{l||cc}
    \hline 
    Galaxy                                              & F25                 & F50                      \\ \hline
    Total baryonic mass [$\times 10^{10} $M$_{\odot}$]  & \multicolumn{2}{c}{5.14}                                                                  \\ \hline
    {\bf Gas Disc (exponential profile)}                               &                      &                              \\
    mass [$\times 10^{10} $M$_{\odot}$]                        & 1.24                   & 2.58                            \\
    characteristic radius [kpc]                         & \multicolumn{2}{c}{2.5}                                                                   \\
    truncation radius [kpc]                          & \multicolumn{2}{c}{5.0}                                                                     \\
    characteristic height [kpc]                      & \multicolumn{2}{c}{0.25}                                                                  \\
    truncation height [kpc]                           & \multicolumn{2}{c}{0.6}                                                                   \\  \hline

    {\bf Stellar disc (exponential profile) }                           &                      &                                    \\
    mass [$\times 10^{10} $M$_{\odot}$]                           & 3.24                   & 1.96                        \\
    characteristic radius [kpc]                         & \multicolumn{2}{c}{2.0}                                                                   \\
    truncation radius [kpc]                          & \multicolumn{2}{c}{4.5}                                                                     \\
    characteristic height [kpc]                      & \multicolumn{2}{c}{0.25}                                                                  \\
    truncation height [kpc]                           & \multicolumn{2}{c}{0.6}                                                                   \\  \hline

    {\bf Bulge (Hernquist profile) }                               &                      &                                 \\
    mass [$\times 10^{10} $M$_{\odot}$]            & \multicolumn{2}{c}{0.58}                 \\
    characteristic height [kpc]                      & \multicolumn{2}{c}{0.25}                                                                  \\
    truncation height [kpc]                           & \multicolumn{2}{c}{0.6}                                                                   \\  \hline
    {\bf Dark Matter halo (Burkert profile) }                      &                      &                                     \\
    mass [$\times 10^{10} $M$_{\odot}$]            & \multicolumn{2}{c}{4.83}                 \\
    characteristic radius [kpc]                      & \multicolumn{2}{c}{8.0}                                                                  \\
    truncation radius [kpc]                           & \multicolumn{2}{c}{14.0}                                                                   \\  \hline

    \end{tabular}
    \end{table}

The initial conditions are built to reach an average \citet{Toomre64} $Q$ parameter of $\sim 0.63$ \citep[close to the critical value for an isothermal disk, ][]{Goldreich65} across the disk, with a stellar\footnote{That is using the velocity dispersion and surface density of the stars.} $Q$ of 1.7 and gaseous $Q$ of 1. We then start the simulations with an adiabatic relaxation phase, during which the gas is maintained at a high temperature, $T = 5 \times 10^5$~K. This phase allows the gas phase to relax within the gravitational potential due to stars and dark matter particles, while not yet forming structures. This phase lasts $\sim$~100~Myr for all runs, which is not counted in the following: we count the time from the end of the relaxation. Maps and radial profiles of the $Q$ parameter and mass surface densities at the end of this phase are shown in Appendix~\ref{App::toomre}. Initializing disks with Q $\sim$ 1.5 - 2 during an orbital time allows preventing numerical biases related to out-of-equilibrium initial conditions \citep[see e.g.][]{Durisen07}. For a discussion of the effect of the relaxation phase on the fragmentation of gas-rich disc we refer the reader to the Section 6.2 of \citet{Faure21} (see Fig.12 and 13), for which we used the same methodology. At the end of the adiabatic relaxation phase, the disk thickness ($\sim$ 1~kpc) is resolved by 30 to 18 cells across the characteristic radius of both the F25 and F50 disks.\\  
  
 Using an adiabatic relaxation phase, a similar cooling model, the same feedback model with a calibration close to our {\it Medium} one and a higher resolution, \citet{Bournaud14} have shown that gas clouds less massive than $10^7$~M$_\odot$ are disrupted within 10 to 20~Myr, validating our feedback models with respect to the lifetime of molecular clouds in nearby galaxies. Such low-mass clumps or clouds are not captured at the resolution of the simulations studied here, and any subsequent claim of long-lived clumps applies only to giant clumps, with masses above $10^{7.5}$~M$_\odot$.

\subsection{Simulation set}

We run simulations with two different gas mass fractions, 25\% and 50\%. They are called F25 and F50 in the following. To do so, we initialize our discs with different gas mass fractions by replacing disc stars by gas. We do not change the other parameters of the simulation, keeping in particular the dark matter and baryonic masses and sizes unchanged. In particular, we keep the same stellar bulge profile, in order to keep the same rotation curve, galactic shear and tidal field for each galaxy of our simulation set. To anchor our results in the existing literature, and to probe the effects of varying the gas fraction while keeping the mass distribution of an existing cosmological simulation,
we use a mass distribution similar to that of the galaxy studied in O17, providing also a similar rotation curve. The associated characteristics are summarized in Table \ref{table::galaxies}. \\

       \begin{figure}
        \centering
         \includegraphics[width=9cm]{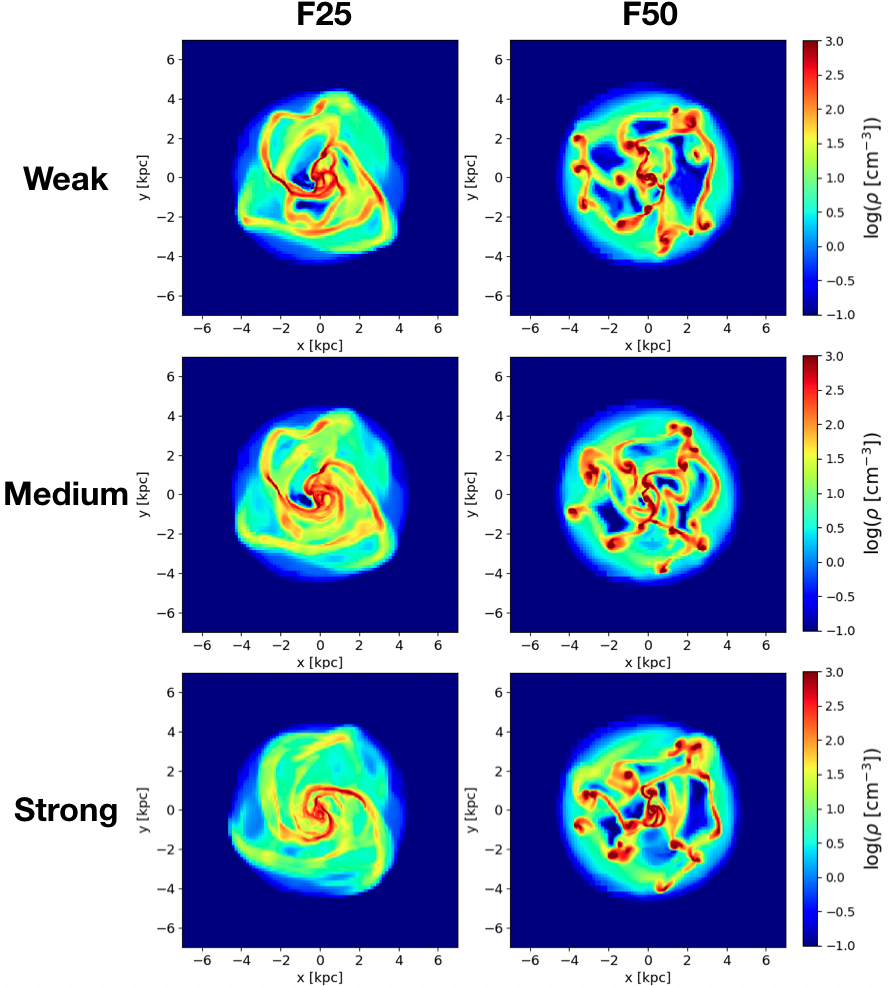}
         \caption{ Gas density map of the simulations after 150\,Myr of simulation
time. The left and right panels show respectively the F25 and F50 models. The rows correspond to different
feedback models as described in Table~\ref{table::galaxies}, from top to bottom: weak,
medium and strong. Each map spans 14~kpc x 14~kpc.
             \label{fig:maps} }
        \end{figure}
        
                \begin{figure}
        \centering
         \includegraphics[width=9cm]{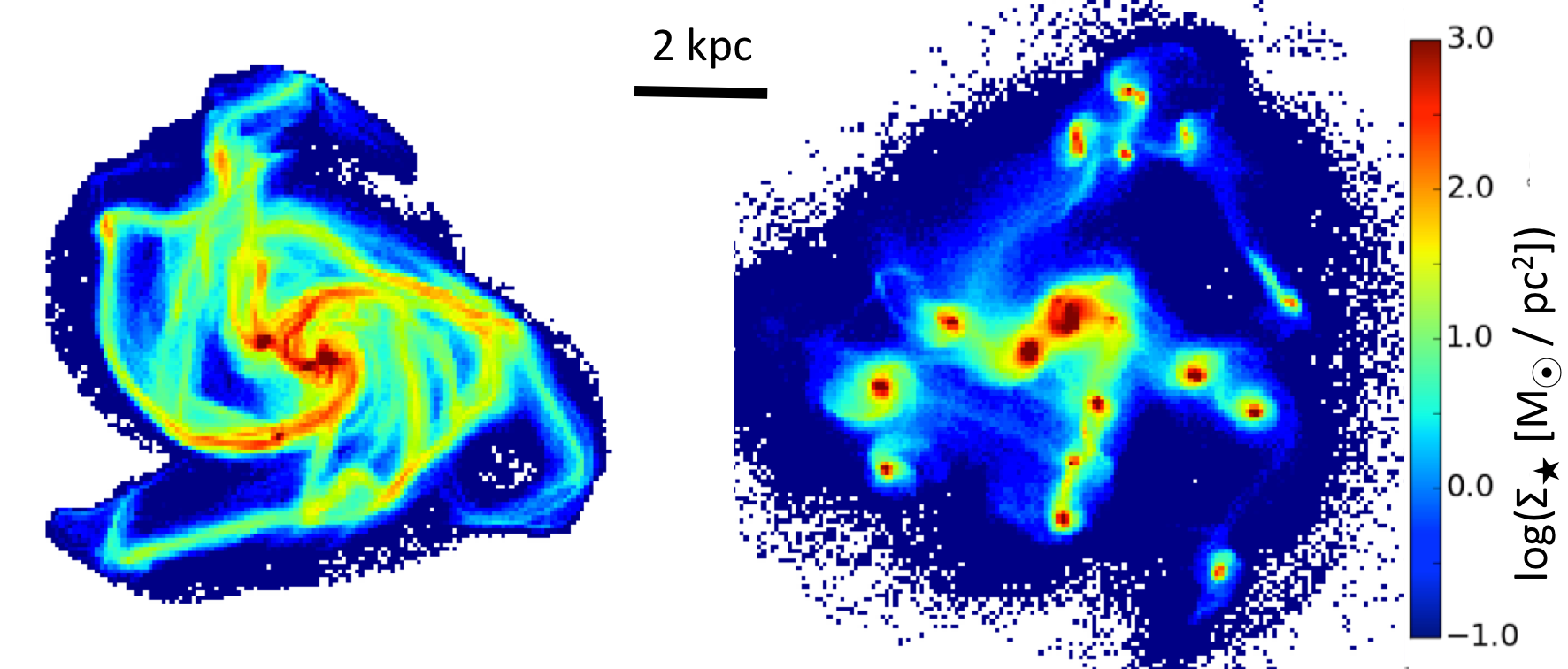}
         \caption{Stellar surface density of stars formed in the simulation for the F25 and F50 simulations and medium feedback after 160\,Myr of simulation time.
             \label{fig:stars} }
         
        \end{figure}

       \begin{figure}
        \centering
         \includegraphics[width=9cm]{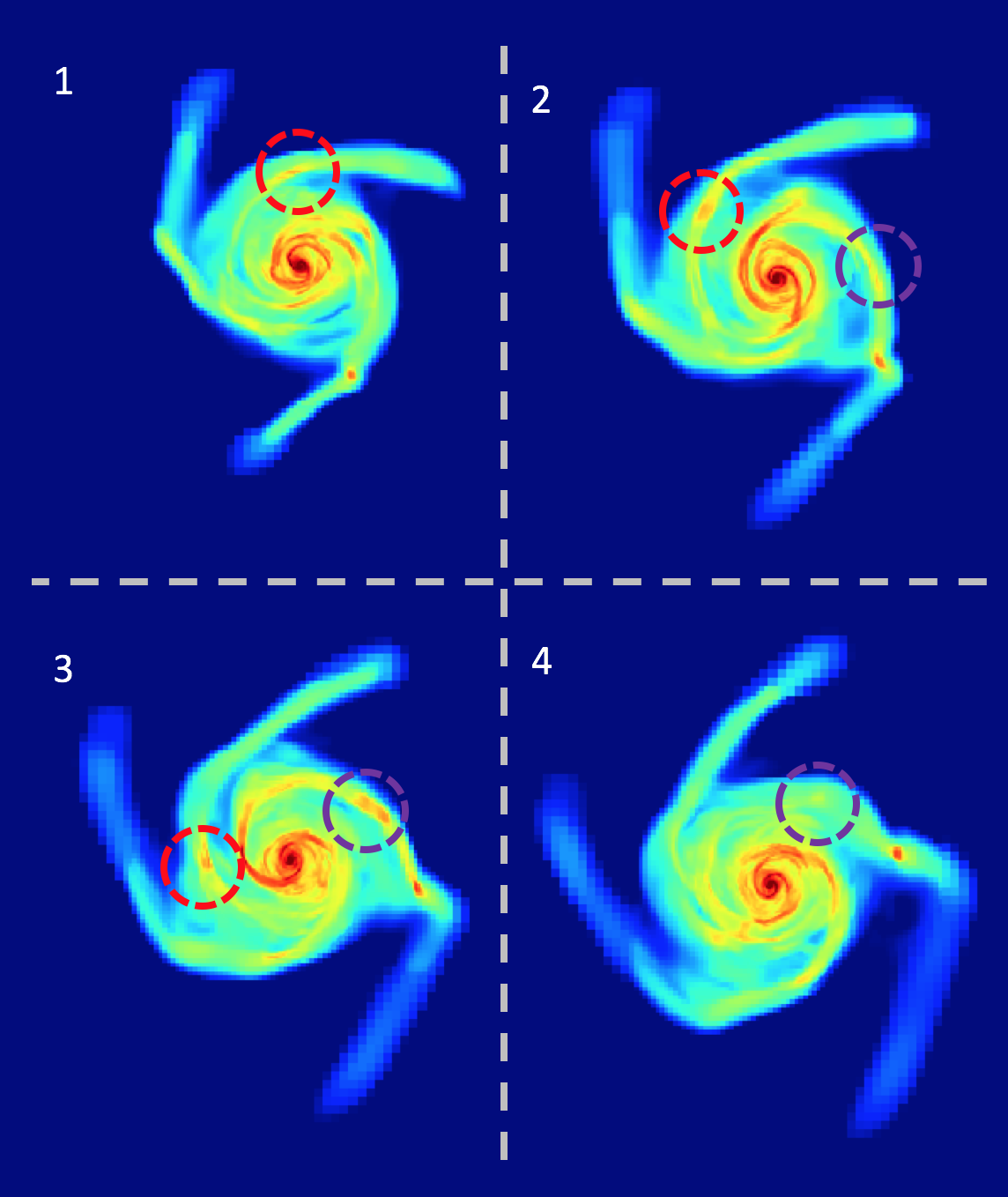}
         \caption{Time sequence of gas density maps from the F25 model run with the medium feedback calibration. The color scale is the same as in Fig.~\ref{fig:maps}. Snapshots are separated by $\sim$\,9\,Myr. The red and purple dashed circles follow the evolution of two gas clumps. 
             \label{fig:F20} }

        \end{figure}

\section{Results}
\label{sec::res}

\subsection{Gas disc evolution}

Figure~\ref{fig:maps} shows the gas density maps of the six simulations, with $f_{\mathrm{gas}}$\,=\,25\% and 50\%, and the three calibrations of feedback, after the same simulation time, 150\,Myr, which corresponds to about one rotation and a half at the stellar half-mass radius. The F25 models show the formation of well-defined spiral structures while the F50 models show the formation of dense ($\rho > 10^{3}$ cm$^{-3}$ ) clumps of gas. Figure~\ref{fig:stars} shows that this clumpy gas distribution indeed translates into a clumpy distribution of the newly-formed stars, while the distribution of young stars in F25 models is spiral-dominated . 

In Fig.~\ref{fig:maps}, one can also see that the stronger models of feedback in the F25  runs make the spiral pattern more diffuse, when they do not seem, at first sight, to affect much the formation of the giant gaseous clump of the F50 runs. In Fig.~\ref{fig:F20} we track two gas clumps in the F25 simulation with medium feedback. We see that they are transient features which remain visible over three snapshots, that is for about $\simeq$\,25\,Myr. This is similar the results from O17 (see their Figure 12).

\subsection{Gas clump evolution}
\label{sec::clump_evol}

       \begin{figure}
        \centering
         \includegraphics[width=9cm]{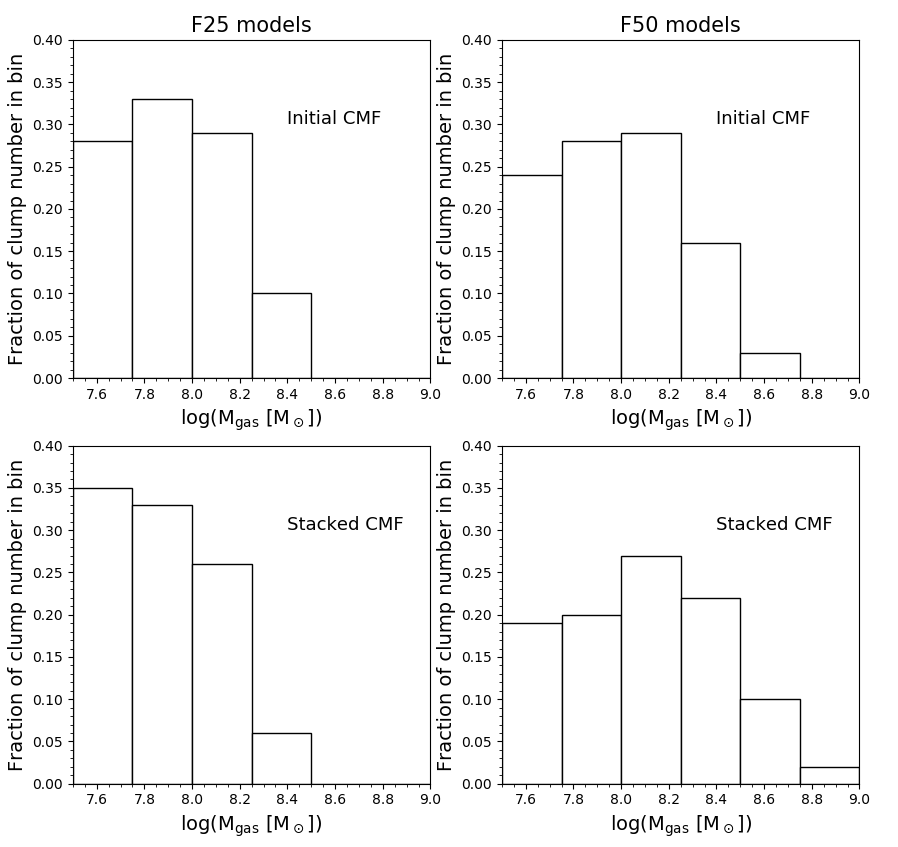}
         \caption{Clump mass function. The left panels shows the F25 models and the right panel shows the F50 models. The top panels show the clump mass function
at the first detection and the bottom panels show the clump mass
function for all detections.
             \label{fig:CMF} }
        \end{figure}

We select clumps on the gas surface density maps. We consider maps of the simulations between 100~Myr and 300~Myr after start, with one snapshot every $\sim 9$~Myr. The maps are smoothed using a Gaussian of FWHM 150~pc and local maximum with a peak above 10~M$_\odot$pc$^{-2}$ are selected as clumps. The extent of each clump is obtained via iterations using concentric circles on the star plus gas column density map, using an incremental value of 50~pc. We stop the iteration when the mass of the previous iteration is close enough to that measured on the background maps (less than a factor 1.33). The background column density is estimated from the mean column density at the galactic radius of the clump, excluding regions less than 0.5\,kpc away from the clump. This technique is similar to observational measures of clump masses (see e.g. \citealt{Guo12}, see also \citealt{Bournaud14}).

The clump mass functions (CMFs thereafter) are shown in Fig.~\ref{fig:CMF}. We show two CMFs for each setup : the first one is the CMF of clumps at first detection, the other one is the stacked CMF of all detected clumps over the simulation (with one snapshot every 9\,Myr). At first detection, clumps in the F50 runs are slightly more massive than in the F25 runs. The distribution peaks at $10^{7.9}$~M$_\odot$ and  $10^{8.1}$~M$_\odot$ for the low- and high-gas mass fraction cases, respectively. The difference between this stacked CMF and the initial CMF shows the mean mass evolution of the clump during their lifetime. For instance, we can see that, in the gas poor case, the bulk of clump moves from  $10^{7.9}$~M$_\odot$ to below  $10^{7.6}$~M$_\odot$. Moreover, the very massive clumps, with mass over  $10^{8.2}$~M$_\odot$, represent around 10\% of the clump abundance in the initial CMF drops to around 5\% on the stacked CMF. This suggests that clump masses tend to follow a global decreasing trend in the gas-poor case simulations.

The opposite trend is observed in the gas-rich case. Comparing the initial CMF and the stacked CMF we see that, if clumps with mass below  $10^{8}$~M$_\odot$ represent 52\% of the initial CMF, their relative abundance drops to below 40\% of the stacked CMF, while the high mass tail of the distribution show an increased abundance. This comparison suggests that on average the mass of clumps increase with time in the gas-rich runs.

One should note that the masses of the gravitationally bound clumps identified on the F25 runs are similar to those in \citet[see their Fig. 4]{Oklopcic17}. The masses obtained in our simulations cannot be readily compared with observed clumps due to a number of observational biases and systematics (such as beam smearing, dust extinction, etc. see \citealt{Faure21} and references therein). 

\begin{figure}
        \centering
         \includegraphics[width=9cm]{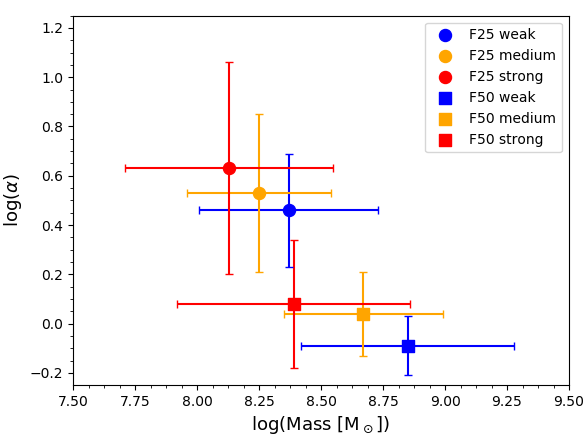}
         \caption{Value of the mean value of the virial $\alpha$ parameter versus total mass of the clumps. The error-bars show the $1-\sigma$ standard deviation of the log of the quantities.
             \label{fig:alpha} }
        \end{figure}

To get a deeper insight on the evolution of these structures, we measure their virial parameter $\alpha$, as defined in O17 by:

\begin{equation}
\alpha =  \frac{5\sigma^{2}R}{\Gr M}
\end{equation}

with $\sigma$ the 1D velocity dispersion along the vertical axis, R and M the radius and mass of the clump measured with the above procedure \citep[see also][]{Bertoldi92}. This parameter enables us to distinguish between bound ($\alpha < 1$) and unbound ($\alpha > 1$) structures. We measure $\alpha$ in our simulations, on clumps of mass between $10^{7.8}$ and $10^{8.2}$ \Msun, to prevent a bias due to the different CMF at the low- and high-mass ends between the F25 and F50 runs. The measurements are done on three snapshots per simulation (at 100, 200 and 300~Myr). The results are summarized in Table~\ref{table:res} and are shown on Fig.~\ref{fig:alpha}.

First, we see that the clumps in the F25 runs are not bound on average, with mean values of $\alpha$ around 3 and around 80\% of the clumps with $\alpha$ above 1.5. These values for $\alpha$ are consistent with the average value found in O17 ($\alpha \simeq 3$, see their Fig.~9). 

On the contrary, clumps of the F50 runs are generally bound with, on average, $\alpha \simeq 1$ (note that the typical tidal field of a massive galaxy disc is expected to significantly stir the clumps, so that very low values $\alpha \ll 1$ are not expected even in the absence of feedback, see \citealt{Elmegreen05}). Increasing the strength of the stellar feedback increases the average $\alpha$ of the clumps in all simulations, but $\alpha$ stays around 1 for the F50 models even for the strongest feedback strength. 

Thus, gas clumps in the F25 models are on average not gravitationally bound, while gas clumps in the F50 models are on average gravitationally bound. We will discuss in Section~\ref{subsec:destrcution} the physical processes responsible for this disruption or survival.

\section{Discussion}
\label{sec::disc}

        \begin{figure}
        \centering
         \includegraphics[width=9cm]{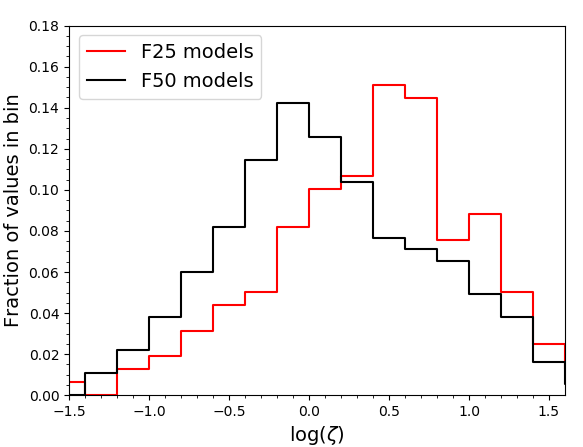}
         \caption{Histogram of the log of the shear parameter $\zeta$ measured for all clump detections, in red for the low gas mass fraction runs and black for the high gas mass fraction case. The histograms are a combination over three snapshots at times $t$=100, 200 and 300~Myr.
             \label{fig:shear} }
        \end{figure}

\begin{table*}
\centering
\caption{Average clump virial parameters, galaxy-wide outflow loading factors, timescales for clump evolution, and inward gas inflow rate for the six simulations. See text for details (Sections~3.2, 4.1, 4.2 and 4.3). The measurements are averaged over three snapshots at times $t$ = 100, 200 and 300~Myr.}
\label{table:res}
\begin{tabular}{l||ccc|ccc}
\hline
          {\bf Galaxy model }                           & \multicolumn{3}{c}{F25}                  & \multicolumn{3}{c}{F50}                      \\
        \textit{\bf Feedback model }               & \textit{weak}     & \textit{medium}  & \textit{strong}   & \textit{weak}       & \textit{medium}   & \textit{strong}    \\ \hline 
 {\bf Average virial parameter $\alpha$ of the clumps } &              &              &                &                &                 &                 \\
 (Section~\ref{sec::clump_evol}) &      2.9        &        3.4      &         4.3       &      0.8          &       1.1          &      1.2           \\ \hline
                        
{\bf Galaxy-wide mass loading factor of galactic winds $\eta$ } &              &               &                &                &                 &                 \\
 (Section~\ref{sec::disc_fbk}) & 0.32    & 1.13 & 3.32 & 0.36  & 0.96 & 3.78                    \\ \hline
{\bf Timescale for... [Myr]  } &              &              &                &                &                 &                 \\
(Section~\ref{subsec:destrcution}) &              &              &                &                &                 &                 \\
Gas removal                      & 75                & 70                & 55                & 520                 & 215                & 195                \\
\textit{(by feedback/stripping)} & \textit{(355/95)} & \textit{(265/95)} & \textit{(195/75)} & \textit{(890/1250)} & \textit{(304/730)} & \textit{(270/700)} \\ \\
Stellar mass loss                & 215               & 200               & 185               & 660                 & 370                & 315                \\ \\
Gas (re-)accretion               & 485               & 405               & 415               & 225                 & 195                & 185                \\ \\
Gas depletion (internal SF)      & 225               & 340               & 360               & 260                 & 275                & 325              \\ \hline 

{\bf Gas inflow rate at 1~kpc  [M$_\odot$ yr$^{-1}$]  }   &              &              &                &                &                 &        \\
(Section~\ref{sec::inflows}) &              &              &                &                &                 &        \\
 
Absolute    & 1.5     & 1.7 & 1.5 & 5.2  & 6.8 & 6.4                    \\ 
Normalized to F50 with medium feedback  & 3.3 & 3.1 & 2.8 & 7.3 & 6.8 & 6.2    \\ \hline

\end{tabular}
\end{table*}

\subsection{Feedback calibration and outflow rates}
\label{sec::disc_fbk}

Stellar feedback calibrations in galaxy simulations being open to debate, one may wonder what one could infer from our results on our feedback calibrations. In Section~\ref{sec::res}, we have seen that the average virial parameters $\alpha$ of the clumps increases with feedback strength. This is expected as our {\it stronger} models were designed to have a stronger impact on the surrounding gas, and decrease the boundedness of star-forming regions.

Another observable related to feedback strength is the mass loading factor $\eta$ of the galaxy winds. It is defined as $\eta = \frac{\mathrm{\dot{M}_w}}{\mathrm{SFR}}$, with $\mathrm{\dot{M}_w}$ the mass outflow rate in M$_\odot$ yr$^{-1}$, and SFR the star formation rate in M$_\odot$ yr$^{-1}$. To measure the mass outflow rate we consider all gas cells that escape a region centered on the galaxy with a velocity higher than the local escape velocity. We chose a box aligned with the disc of height $\pm 5$\,kpc and which span the whole box in the other two directions. Choosing a cylinder of same height and of radius 20\,kpc changes the results by only a few percents. The values of the mass loading factors are given in Table~\ref{table::bffb}.

The loading factor $\eta$ is largely independent on the gas fraction and similar for F25 and F50 runs for any given feedback calibration. The average values of $\eta$ are about 0.3, 1 and 3.5 for the {\it Weak}, {\it Medium} and {\it Strong} feedback calibrations, respectively. Observed samples of galaxies in the same mass range at $z$ $\simeq$1-2 typically have a loading factor of the order of unity when probed in front of background quasars, with extreme values ranging from $\sim$ 0.1 to $\sim$10 \citep[e.g.,][]{Schroetter19}. Direct observations of outflowing gas yield similar conclusions, with average loading factors rate somewhat below unity or at most of the order or unity \citep[e.g.,][]{Forster19}. Our three feedback calibrations roughly encompass the observed plausible range, hence deserving to be labelled {\it Weak}, {\it Medium} and {\it Strong}, respectively. In particular, our {\it Strong} feedback calibration appears to represent the highest plausible bound in terms of galaxy-scale outflow rates. In contrast, it is interesting to note that cosmological simulations generally\footnote{
     Two noticeable exceptions are the Horizon-AGN simulation \citep{Dubois16, Beckmann17} with mass loading factors of the order of unity \citep{Chabanier20}, as well as the VELA simulation series \citep{Ceverino14,Zolotov15}. Horizon-AGN simulation is a large volume simulation that cannot resolve giant clump formation, but the NewHorizon simulation that employs a largely similar physical model does produce irregular and clumpy disks at $z$ $\simeq$ 2 \citep[][, Fig.~4]{Dubois20} ; VELA zoom-in simulations do resolve giant clump formation and predict long-lived clumps \citep{Mandelker17}.
} have higher outflow rates, of the order of 10 or above for star-forming galaxies in the same mass range at $z$=1-2 (e.g., \citealt{Muratov15} for the FIRE simulation series and \citealt{Nelson19} for the Illustris-TNG model). The underlying feedback models provide realistic total galaxy masses and stellar-to-halo mass relation at z=0 \citep[see e.g.][]{Pillepich19} but these strong outflows may relate to the low gas fractions in these simulations (see Introduction) compared to $z$ $\simeq$ 2 star-forming galaxies, resulting in their relative stability against giant clump formation. 

One may wonder whether our feedback calibration may lead to the observed stellar-to-halo mass relation at z=0. We have tried three different calibrations, spanning a wide range in outflow mass loading factors, which yielded similar results. Thus, our results infer that the stellar feedback only plays a second order role on the fragmentation of the disc. This low effect of stellar feedback calibrations may seem to be in contradiction with other studies (see e.g. \citealt{Tamburello15, Mandelker17}, but see \citealt{Moody14}). For instance \citet{Mayer16} find that their implementation of super-bubble feedback drastically reduced the mass of clumps, despite a high gas mass fraction in the disk ($\sim 50\%$). It should be noted that this implementation implies mass loading factors above 5 for 1<z<3 galaxies \citep[see Fig. 12]{Keller15}, which makes it a stronger calibration than even our Strong calibration. \citet{Tamburello15} used blast wave feedback, which induces mass loading factors around 1 \citep{Keller15}, but had no long-lived clumps in their isolated gas-rich disks. One should note that their galaxies have a larger size and that their total mass surface density at one disk length scale at the end of relaxation is significantly lower than in our F50 models (see their Figure 6 and Figure~\ref{fig::A_surfaced}). Ultimately, as the onset the clumpy phase depends on the whole structure of the galaxy, such as its average surface density and compactness, it is difficult to compare studies using other galaxy models.\\

We recall the reader that a similar cooling model and a calibration close to our {\it Medium} feedback model, while producing long-lived giant clumps for a high enough gas fraction, produces short-lived clumps for low masses below $\sim$\,$10^7$~M$_\odot$ \citep{Bournaud14}.

\subsection{Origin of clump destruction}
\label{subsec:destrcution}

To understand which processes make these clumps lose or gain mass we do a detailed analysis of the gas and star behavior around the clump. For each cell of gas, we compute the relative velocity with respect to the center of mass of the clump. If a linear extrapolation from this velocity on a short time-step (1~Myr) causes the gas to cross the border of the clump inwards, it is counted as gas accretion, if it crosses outwards it is counted as gas removal. We furthermore separate gas removal by {\it feedback} and by tidal {\it stripping}, assuming roughly that gas removed by feedback has been warmed-up above $10^{4}$~K (in HII regions if not through SN explosions) while tidally-stripped gas is assumed to remain colder. The same method is applied to stars inside the clump size. We also account for gas depletion by star formation by measuring the instantaneous star formation rate in the clump. We then divide each of these values by the mass of the clump to get the timescale for gas and star loss and accretion. The results are summed up in Table~\ref{table:res}.

This table quantifies what is seen in the previous figures, namely that gas removal of gaseous clumps is very efficient in the low gas mass fraction case. It should be noted that this gas removal is mainly due to stripping -- induced by shear and tidal effects -- while in the gas-rich case, the gas removal is mainly due to feedback. \\

One should note that in the low gas mass fraction cases, gas clumps form in spiral arms which are regions of maximal shear (e.g., \citealt{Renaud13}, Fig.~13, \citealt{Emsellem15}, see also \citealt{Brucy20}). We measure the shear $\zeta$ of each clump via the following procedure. We measure the mean tangential velocity of cells in a region of diameter 50\,pc in two regions centered on the intersection of the boundary of the clump and the line between the clump and the galaxy center. The inner (resp. outer) value is noted $V_1$ (resp. $V_2$). We then measure the rotational velocity at the radial distance $R$ of the clump to the galaxy center, $V(R)$. The shear velocity is then defined as $V_s = \frac{V_2 - V_1}{2} - V(R)$. Last, we measure the rotational velocity due to the own gravity of the clump of mass $M$ at its half-mass radius $R_{1/2}$, given by $V_\mathrm{clump} = \sqrt{\frac{0.5 G M}{R_{1/2}}}$. The shear parameter $\zeta$ is then defined as $\zeta = V_s / V_\mathrm{clump}$. It is a competition between rotational motions of gas due to the clump, and differential rotational motions due to the galactic disc motions. Shear dominates the internal motion for $\zeta$\,>\,1.

The obtained $\zeta$ values for clumps in the low- and high-gas mass fraction runs are presented in Fig.~\ref{fig:shear} combined over three snapshots at times $t$=100, 200 and 300~Myr. We can see that, on average, gas clumps in the low gas mass fraction case are located in regions of higher shear than in the high gas mass fraction case: indeed in the low gas fraction models, clumps form only in spiral arms (Fig.~\ref{fig:maps}) while in high gas fraction cases, they can form throughout the disk. We remind that the total mass distribution is the same for our F25 and F50 models, hence the azimuthally-average shear profile is the same in the two cases: preferential location of clumps in spiral arms (other regions being stable against clump formation) in the gas-poor cases may thus explain the higher shear undergone by clumps in these cases. 

To verify that the destruction of the giant clumps is mainly due to shear in the F25 runs, we compare three sub-samples of clumps, matched in a given physical quantity, for the F25 and F50 runs with {\it Medium} feedback, over three snapshots (100, 200 and 300~Myr):

\begin{itemize}
    \item A mass-selected sub-sample: we select only clumps that have a total baryonic mass between $10^{8.2}$ and $10^{8.6}$~M$_\odot$ (resulting in a mean mass of $10^{8.32}$ and $10^{8.47}$ for the F25 and F50 runs, respectively). These samples comprise 21 and 26 clumps for  the F25 and F50 runs, respectively. They have half-mass radius of 292$\pm$87~pc and 268$\pm$103~pc for the F25 and F50 runs, respectively. The removal timescale of the gas by feedback/stripping, as defined at the begenning of Section~\ref{subsec:destrcution}, is 257/98~Myr and 294/692~Myr for the F25 and F50 runs, respectively.
    \item A mass-matched sub-sample: for each clump in a F50 run, we randomly match a F25 clump which has a baryonic mass within 0.1~dex, if any. If there is no such clump in the F25 dataset, we reject the F50 clump. We limit this sub-sample to clumps with baryonic mass between $10^8$ and $10^9$~M$_\odot$. These sub-samples comprise 11 clumps for both the F25 and F50 runs, respectively. The mean baryonic mass of clumps in this sample is $10^{8.39}$~M$_\odot$ for both F25 and F50 runs. The removal timescale of the gas by feedback/stripping is 283/108~Myr and 315/739~Myr for the F25 and F50 runs, respectively.
    \item A gas fraction-matched sub-sample: we select all clumps in the F25 and F50 runs with a gas fraction between 0.3 and 0.4 and a baryonic mass between $10^{8}$ and $10^{9}$~M$_\odot$. The gas fraction is measured within the half-mass radius of each clump. These sub-samples comprise 13 and 18 clumps for  the F25 and F50 runs, respectively. The removal timescale of the gas by feedback/stripping is 276/112~Myr and 289/608~Myr for the F25 and F50 runs, respectively.
\end{itemize}

    \begin{table}
    \centering
    \caption{Timescale for gas loss due to feedback and stripping for different sub-samples of clumps in the F25 and F50 runs with medium feedback.}
    \label{table::shear}
    \begin{tabular}{lcc}
    \hline 
    Sub-sample & \multicolumn{2}{l}{Timescale for gas removal}   \\
                 & \multicolumn{2}{l}{by feedback/stripping [Myr]}   \\\hline
                 & F25  & F50   \\
                \hline 
     No match & 265/95 & 304/730   \\          
     \\
    Mass-selected & 257/98 & 294/692   \\  
    Mass-matched &  283/108 & 315/739   \\ 
    Gas fraction-matched &  276/112 &  289/608  \\ 
 
    \hline
    \end{tabular}
    \end{table} 

These results are summarized in Table~\ref{table::shear}. We see that for clumps with similar mass, similar radius or similar gas fraction the gas removal via shear is much stronger than for F50 clumps, while the gas expulsion timescale via feedback remains similar.

Thus we conclude that destruction by shear rather than by feedback in the low gas fraction cases explain why the clump lifetime is rather independent on the feedback calibration (Table~\ref{table:res}).

\subsection{Gas mass fraction and gas inflows}
\label{sec::inflows}

In the previous sections, we have seen that the high gas mass fractions trigger the formation of long-lived clumps, which resist both shear and feedback. High gas fraction discs also have strong nuclear gas inflows, due to clump inward migration \citep{Noguchi99, Elmegreen05, Bournaud07} and the instability-driven inflow (giant clumps also torque diffuse gas inward, \citealt{Bournaud12, Dekel13}). The energy released by these inflows is though to fuel the elevated turbulence and self-regulate the Toomre parameter of gas-rich discs \citep[see e.g.][]{Dekel09}. 
We measure the absolute gas inflow rate within the central kpc, using the methods as for the clumps. The results are shown in the bottom part of Table~\ref{table::bffb}. We see that the F25 and F50 simulations have absolute gas inflow rates respectively between 1.5 and 1.7~M$_\odot$yr$^{-1}$ and between 5.2 and 6.8~M$_\odot$yr$^{-1}$.

However, the mass of gas in the galaxy at a given time is not the same for each run. It depends on the galaxy star formation history and thus on the initial gas mass fraction and feedback calibration. In the last line we re-normalize the inflow rate by multiplying them  by $\mathrm{M}_\mathrm{gas, i} / \mathrm{M}_\mathrm{gas, F50-medium}$ for each run $i$. After this normalization, the inflow rates of the F25 runs remain around a factor two lower than the F50 runs, namely 2.8-3.3~M$_\odot$yr$^{-1}$ against  6.2-7.3~M$_\odot$yr$^{-1}$. This shows that the difference in disc instability between the F25 and F50 runs is not only on the galaxy morphology but would affect the motion of the gas in the galactic discs as a whole. On a longer timescale, these instability driven inflows would impact the formation of the bulge and the radial profile of the galaxy: changing the gas fraction of a galaxy from 25\% to 50\% will hence change the rate of bulge growth and potential nuclear gas fuelling that can result from giant clump formation and evolution \citep[see review by ][]{Bournaud16}.

\subsection{Effects of the environment}
\label{subsec:env}

Our simulations reproduce the mass distribution of the galaxy studied in O17. But unlike the O17 cosmological simulation, they start from an idealized disc and do not account for the cosmological environment, such as gas accretion or galaxy interactions and mergers.

Infalling material would not does not settle immediately into a thin disc but would tend to stir up the galaxy and increase the disc scale-height and turbulence (see e.g. \citealt{Elmegreen10, Gabor14}, but see \citealt{Hopkins13}). In \citet{Fensch17} we have shown that the turbulence induced by the gas-rich disc instability is already high, such that the release of gravitational energy from an equal-mass merger of gas dominated galaxies does not significantly increase the gas disc turbulence. We argue that including  a cosmological background should therefore not significantly impact the development of these disk instabilities and thus not modify our results.
If anything, rare major mergers could de-stabilize low-gas fraction disks into low-mass clumps \citep{Teyssier10, Renaud15}. The effect of major mergers at high gas fractions (30\% to 50\%) is milder, with a limited impact on the star formation rate and clumpiness \citep{Fensch17, Calabro19}.

\section{Conclusion}
\label{sec::conc}

We have studied the difference in disc stability and evolution between galaxies with a medium and a high gas mass fraction, namely 25\% and 50\% of the baryonic mass. These values are motivated by the typical values of gas fraction in cosmological simulations ($\sim 25\%$) and in observations ($\sim 50\%$).

We performed idealised disc simulations with these two gas mass fraction and similar total mass distribution, and study the evolution of the gas clumps which form in them. We show that for models with 25\% of gas the clumps are unbound and are destroyed in less than 100~Myr, whereas clumps in models with 50\% of gas are gravitationally bound. By using three different feedback calibrations, we show that our results are robust against calibrations which yield too weak or too strong galaxy-wide gas outflows. At low gas fraction, clumps are destroyed quickly ($\leq$50\,Myr) even with weak feedback, while at high gas fraction, massive-enough clumps are long-lived ($\geq$100\,Myr) even with strong feedback. We show that clumps in the lower gas fraction models undergo strong shear, because at low gas fraction, big clumps form only in spiral arms (other regions being stable), and spiral arms are regions with high shear, eventually causing the destruction of clumps independently of feedback parameters. Our models show that, on $\sim$$10^8$\,yr timescales, the gas fraction has a stronger effect on clump masses, boundedness and lifetimes, than the stellar feedback calibration, at least for a range of stellar feedback parameters consistent with typical gas outflow rates observed for high-redshift star-forming galaxies.

These results could explain why cosmological simulations that have relatively low gas fractions at $z\sim2$ lack long-lived giant clumps, as opposed to simulations with higher gas fractions (see Introduction). The details of feedback modelling in these simulations seems to have a weaker impact on clump evolution on $\sim$$10^8$\,yr timescales. On longer cosmological timescales, the evolution of the gas fraction can be influenced by feedback and outflows, but also results form many other processes such as gas accretion flows, cooling, shocks, etc.

 We show that another effect of these disc instabilities is that clump migration fuels strong gas nuclear inflows. This shows that the disc instabilities caused by the high gas fraction is not likely to be only summarized by a morphology difference at a given time, but on the whole evolution of the structure of the galaxies. Given the large impact of violent disc instabilities on the morphology of galaxies and their evolution, future cosmological simulations should particularly aim at maintaining realistic gas mass fractions along with realistic gas outflow rates in star-forming galaxies across cosmic times.

\section*{Acknowledgements}
 We thank the referee for their careful reading and comments that helped improve the paper. We acknowledge numerous discussions on disc stability, clumps, shear and turbulence with Andi Burkert, Avishai Dekel, Bruce Elmegreen, Reinhard Genzel, Nir Mandelker and Florent Renaud. We also acknowledge many discussions on outflow rates in cosmological simulations with Yohan Dubois and Ricarda Beckmann, and in observations with Nicolas Bouch\'e. We are grateful to Pierre-Alain Duc for his support to this project. This work was performed using HPC resources from GENCI at TGCC and CINES (grants A0050402192, A0070402192 and A0090402192).
 FB acknowledges support from the ANR through the 3DGasFlows project (ANR-17-CE31-0017).

\section*{Data Availability Statement}
The data underlying this article will be shared on reasonable request to the corresponding author.

\bibliographystyle{mnras}
\bibliography{example} 

\begin{thebibliography}{}
\makeatletter
\relax
\def\mn@urlcharsother{\let\do\@makeother \do\$\do\&\do\#\do\^\do\_\do\%\do\~}
\def\mn@doi{\begingroup\mn@urlcharsother \@ifnextchar [ {\mn@doi@}
  {\mn@doi@[]}}
\def\mn@doi@[#1]#2{\def\@tempa{#1}\ifx\@tempa\@empty \href
  {http://dx.doi.org/#2} {doi:#2}\else \href {http://dx.doi.org/#2} {#1}\fi
  \endgroup}
\def\mn@eprint#1#2{\mn@eprint@#1:#2::\@nil}
\def\mn@eprint@arXiv#1{\href {http://arxiv.org/abs/#1} {{\tt arXiv:#1}}}
\def\mn@eprint@dblp#1{\href {http://dblp.uni-trier.de/rec/bibtex/#1.xml}
  {dblp:#1}}
\def\mn@eprint@#1:#2:#3:#4\@nil{\def\@tempa {#1}\def\@tempb {#2}\def\@tempc
  {#3}\ifx \@tempc \@empty \let \@tempc \@tempb \let \@tempb \@tempa \fi \ifx
  \@tempb \@empty \def\@tempb {arXiv}\fi \@ifundefined
  {mn@eprint@\@tempb}{\@tempb:\@tempc}{\expandafter \expandafter \csname
  mn@eprint@\@tempb\endcsname \expandafter{\@tempc}}}

\bibitem[\protect\citeauthoryear{Agertz, Lake, Teyssier, Moore, Mayer  \&
  Romeo}{Agertz et~al.}{2009}]{Agertz09}
Agertz O.,  Lake G.,  Teyssier R.,  Moore B.,  Mayer L.,   Romeo A.~B.,  2009,
  \mn@doi [Monthly Notices of the Royal Astronomical Society]
  {10.1111/j.1365-2966.2008.14043.x}, 392, 294–308

\bibitem[\protect\citeauthoryear{{Beckmann} et~al.,}{{Beckmann}
  et~al.}{2017}]{Beckmann17}
{Beckmann} R.~S.,  et~al., 2017, \mn@doi [\mnras] {10.1093/mnras/stx1831},
  \href {https://ui.adsabs.harvard.edu/abs/2017MNRAS.472..949B} {472, 949}

\bibitem[\protect\citeauthoryear{{Behrendt}, {Burkert}  \&
  {Schartmann}}{{Behrendt} et~al.}{2016}]{Behrendt16}
{Behrendt} M.,  {Burkert} A.,   {Schartmann} M.,  2016, \mn@doi [\apjl]
  {10.3847/2041-8205/819/1/L2}, \href
  {https://ui.adsabs.harvard.edu/abs/2016ApJ...819L...2B} {819, L2}

\bibitem[\protect\citeauthoryear{{Bertoldi} \& {McKee}}{{Bertoldi} \&
  {McKee}}{1992}]{Bertoldi92}
{Bertoldi} F.,  {McKee} C.~F.,  1992, \mn@doi [\apj] {10.1086/171638}, \href
  {http://adsabs.harvard.edu/abs/1992ApJ...395..140B} {395, 140}

\bibitem[\protect\citeauthoryear{{Bournaud}}{{Bournaud}}{2016}]{Bournaud16}
{Bournaud} F.,  2016, {Bulge Growth Through Disc Instabilities in High-Redshift
  Galaxies}.
p.~355, \mn@doi{10.1007/978-3-319-19378-6_13}

\bibitem[\protect\citeauthoryear{{Bournaud}, {Elmegreen}  \&
  {Elmegreen}}{{Bournaud} et~al.}{2007}]{Bournaud07}
{Bournaud} F.,  {Elmegreen} B.~G.,   {Elmegreen} D.~M.,  2007, \mn@doi [\apj]
  {10.1086/522077}, \href
  {https://ui.adsabs.harvard.edu/abs/2007ApJ...670..237B} {670, 237}

\bibitem[\protect\citeauthoryear{{Bournaud} et~al.,}{{Bournaud}
  et~al.}{2011a}]{Bournaud11}
{Bournaud} F.,  et~al., 2011a, \mn@doi [\apj] {10.1088/0004-637X/730/1/4},
  \href {http://cdsads.u-strasbg.fr/abs/2011ApJ...730....4B} {730, 4}

\bibitem[\protect\citeauthoryear{{Bournaud}, {Dekel}, {Teyssier}, {Cacciato},
  {Daddi}, {Juneau}  \& {Shankar}}{{Bournaud} et~al.}{2011b}]{Bournaud12}
{Bournaud} F.,  {Dekel} A.,  {Teyssier} R.,  {Cacciato} M.,  {Daddi} E.,
  {Juneau} S.,   {Shankar} F.,  2011b, \mn@doi [\apjl]
  {10.1088/2041-8205/741/2/L33}, \href
  {https://ui.adsabs.harvard.edu/abs/2011ApJ...741L..33B} {741, L33}

\bibitem[\protect\citeauthoryear{{Bournaud} et~al.,}{{Bournaud}
  et~al.}{2014}]{Bournaud14}
{Bournaud} F.,  et~al., 2014, \mn@doi [\apj] {10.1088/0004-637X/780/1/57},
  \href {http://cdsads.u-strasbg.fr/abs/2014ApJ...780...57B} {780, 57}

\bibitem[\protect\citeauthoryear{{Brucy}, {Hennebelle}, {Bournaud}  \&
  {Colling}}{{Brucy} et~al.}{2020}]{Brucy20}
{Brucy} N.,  {Hennebelle} P.,  {Bournaud} F.,   {Colling} C.,  2020, \mn@doi
  [\apjl] {10.3847/2041-8213/ab9830}, \href
  {https://ui.adsabs.harvard.edu/abs/2020ApJ...896L..34B} {896, L34}

\bibitem[\protect\citeauthoryear{{Buck}, {Macci{\`o}}, {Obreja}, {Dutton},
  {Dom{\'\i}nguez-Tenreiro}  \& {Granato}}{{Buck} et~al.}{2017}]{Buck17}
{Buck} T.,  {Macci{\`o}} A.~V.,  {Obreja} A.,  {Dutton} A.~A.,
  {Dom{\'\i}nguez-Tenreiro} R.,   {Granato} G.~L.,  2017, \mn@doi [\mnras]
  {10.1093/mnras/stx685}, \href
  {https://ui.adsabs.harvard.edu/abs/2017MNRAS.468.3628B} {468, 3628}

\bibitem[\protect\citeauthoryear{{Calabr{\`o}} et~al.,}{{Calabr{\`o}}
  et~al.}{2019}]{Calabro19}
{Calabr{\`o}} A.,  et~al., 2019, \mn@doi [\aap] {10.1051/0004-6361/201935778},
  \href {https://ui.adsabs.harvard.edu/abs/2019A&A...632A..98C} {632, A98}

\bibitem[\protect\citeauthoryear{Cava, Schaerer, Richard, Pérez-González,
  Dessauges-Zavadsky, Mayer  \& Tamburello}{Cava et~al.}{2018}]{Cava18}
Cava A.,  Schaerer D.,  Richard J.,  Pérez-González P.~G.,
  Dessauges-Zavadsky M.,  Mayer L.,   Tamburello V.,  2018, \mn@doi [Nature
  Astronomy] {10.1038/s41550-017-0295-x}, 2, 76–82

\bibitem[\protect\citeauthoryear{{Ceverino}, {Klypin}, {Klimek},
  {Trujillo-Gomez}, {Churchill}, {Primack}  \& {Dekel}}{{Ceverino}
  et~al.}{2014}]{Ceverino14}
{Ceverino} D.,  {Klypin} A.,  {Klimek} E.~S.,  {Trujillo-Gomez} S.,
  {Churchill} C.~W.,  {Primack} J.,   {Dekel} A.,  2014, \mn@doi [\mnras]
  {10.1093/mnras/stu956}, \href
  {http://adsabs.harvard.edu/abs/2014MNRAS.442.1545C} {442, 1545}

\bibitem[\protect\citeauthoryear{{Chabanier}, {Bournaud}, {Dubois},
  {Palanque-Delabrouille}, {Y{\`e}che}, {Armengaud}, {Peirani}  \&
  {Beckmann}}{{Chabanier} et~al.}{2020}]{Chabanier20}
{Chabanier} S.,  {Bournaud} F.,  {Dubois} Y.,  {Palanque-Delabrouille} N.,
  {Y{\`e}che} C.,  {Armengaud} E.,  {Peirani} S.,   {Beckmann} R.,  2020,
  \mn@doi [\mnras] {10.1093/mnras/staa1242}, \href
  {https://ui.adsabs.harvard.edu/abs/2020MNRAS.495.1825C} {495, 1825}

\bibitem[\protect\citeauthoryear{{Cowie}, {Hu}  \& {Songaila}}{{Cowie}
  et~al.}{1995}]{Cowie95}
{Cowie} L.~L.,  {Hu} E.~M.,   {Songaila} A.,  1995, \mn@doi [\nat]
  {10.1038/377603a0}, \href
  {https://ui.adsabs.harvard.edu/abs/1995Natur.377..603C} {377, 603}

\bibitem[\protect\citeauthoryear{{Daddi} et~al.,}{{Daddi}
  et~al.}{2010a}]{Daddi10}
{Daddi} E.,  et~al., 2010a, \mn@doi [\apj] {10.1088/0004-637X/713/1/686}, \href
  {https://ui.adsabs.harvard.edu/abs/2010ApJ...713..686D} {713, 686}

\bibitem[\protect\citeauthoryear{{Daddi} et~al.,}{{Daddi}
  et~al.}{2010b}]{Daddi10a}
{Daddi} E.,  et~al., 2010b, \mn@doi [\apjl] {10.1088/2041-8205/714/1/L118},
  \href {http://cdsads.u-strasbg.fr/abs/2010ApJ...714L.118D} {714, L118}

\bibitem[\protect\citeauthoryear{{Dekel} \& {Krumholz}}{{Dekel} \&
  {Krumholz}}{2013}]{Dekel13}
{Dekel} A.,  {Krumholz} M.~R.,  2013, \mn@doi [\mnras] {10.1093/mnras/stt480},
  \href {https://ui.adsabs.harvard.edu/abs/2013MNRAS.432..455D} {432, 455}

\bibitem[\protect\citeauthoryear{Dekel \& Mandelker}{Dekel \&
  Mandelker}{2014}]{Dekel14}
Dekel A.,  Mandelker N.,  2014, \mn@doi [Monthly Notices of the Royal
  Astronomical Society] {10.1093/mnras/stu1427}, 444, 2071–2084

\bibitem[\protect\citeauthoryear{Dekel, Sari  \& Ceverino}{Dekel
  et~al.}{2009}]{Dekel09}
Dekel A.,  Sari R.,   Ceverino D.,  2009, \mn@doi [The Astrophysical Journal]
  {10.1088/0004-637X/703/1/785}, 703, 785–801

\bibitem[\protect\citeauthoryear{{Dessauges-Zavadsky}
  et~al.,}{{Dessauges-Zavadsky} et~al.}{2019}]{DEssauges-Zavadsky19}
{Dessauges-Zavadsky} M.,  et~al., 2019, \mn@doi [Nature Astronomy]
  {10.1038/s41550-019-0874-0}, \href
  {https://ui.adsabs.harvard.edu/abs/2019NatAs...3.1115D} {3, 1115}

\bibitem[\protect\citeauthoryear{{Dubois} \& {Teyssier}}{{Dubois} \&
  {Teyssier}}{2008}]{Dubois08}
{Dubois} Y.,  {Teyssier} R.,  2008, \mn@doi [\aap]
  {10.1051/0004-6361:20078326}, \href
  {http://cdsads.u-strasbg.fr/abs/2008A%26A...477...79D} {477, 79}

\bibitem[\protect\citeauthoryear{{Dubois}, {Peirani}, {Pichon}, {Devriendt},
  {Gavazzi}, {Welker}  \& {Volonteri}}{{Dubois} et~al.}{2016}]{Dubois16}
{Dubois} Y.,  {Peirani} S.,  {Pichon} C.,  {Devriendt} J.,  {Gavazzi} R.,
  {Welker} C.,   {Volonteri} M.,  2016, \mn@doi [\mnras]
  {10.1093/mnras/stw2265}, \href
  {https://ui.adsabs.harvard.edu/abs/2016MNRAS.463.3948D} {463, 3948}

\bibitem[\protect\citeauthoryear{{Dubois} et~al.,}{{Dubois}
  et~al.}{2020}]{Dubois20}
{Dubois} Y.,  et~al., 2020, arXiv e-prints, \href
  {https://ui.adsabs.harvard.edu/abs/2020arXiv200910578D} {p. arXiv:2009.10578}

\bibitem[\protect\citeauthoryear{{Durisen}, {Boss}, {Mayer}, {Nelson}, {Quinn}
  \& {Rice}}{{Durisen} et~al.}{2007}]{Durisen07}
{Durisen} R.~H.,  {Boss} A.~P.,  {Mayer} L.,  {Nelson} A.~F.,  {Quinn} T.,
  {Rice} W.~K.~M.,  2007, in {Reipurth} B.,  {Jewitt} D.,   {Keil} K.,  eds,
  Protostars and Planets V. p.~607 (\mn@eprint {arXiv} {astro-ph/0603179})

\bibitem[\protect\citeauthoryear{Elmegreen \& Burkert}{Elmegreen \&
  Burkert}{2010}]{Elmegreen10}
Elmegreen B.~G.,  Burkert A.,  2010, \mn@doi [The Astrophysical Journal]
  {10.1088/0004-637X/712/1/294}, 712, 294–302

\bibitem[\protect\citeauthoryear{{Elmegreen}, {Elmegreen}, {Vollbach}, {Foster}
   \& {Ferguson}}{{Elmegreen} et~al.}{2005}]{Elmegreen05}
{Elmegreen} B.~G.,  {Elmegreen} D.~M.,  {Vollbach} D.~R.,  {Foster} E.~R.,
  {Ferguson} T.~E.,  2005, \mn@doi [\apj] {10.1086/496952}, \href
  {https://ui.adsabs.harvard.edu/abs/2005ApJ...634..101E} {634, 101}

\bibitem[\protect\citeauthoryear{{Elmegreen}, {Elmegreen}, {Ravindranath}  \&
  {Coe}}{{Elmegreen} et~al.}{2007}]{Elmegreen07}
{Elmegreen} D.~M.,  {Elmegreen} B.~G.,  {Ravindranath} S.,   {Coe} D.~A.,
  2007, \mn@doi [\apj] {10.1086/511667}, \href
  {https://ui.adsabs.harvard.edu/abs/2007ApJ...658..763E} {658, 763}

\bibitem[\protect\citeauthoryear{{Elmegreen}, {Bournaud}  \&
  {Elmegreen}}{{Elmegreen} et~al.}{2008}]{Elmegreen08}
{Elmegreen} B.~G.,  {Bournaud} F.,   {Elmegreen} D.~M.,  2008, \mn@doi [\apj]
  {10.1086/592190}, \href
  {https://ui.adsabs.harvard.edu/abs/2008ApJ...688...67E} {688, 67}

\bibitem[\protect\citeauthoryear{{Emsellem}, {Renaud}, {Bournaud}, {Elmegreen},
  {Combes}  \& {Gabor}}{{Emsellem} et~al.}{2015}]{Emsellem15}
{Emsellem} E.,  {Renaud} F.,  {Bournaud} F.,  {Elmegreen} B.,  {Combes} F.,
  {Gabor} J.~M.,  2015, \mn@doi [\mnras] {10.1093/mnras/stu2209}, \href
  {https://ui.adsabs.harvard.edu/abs/2015MNRAS.446.2468E} {446, 2468}

\bibitem[\protect\citeauthoryear{{Faure}, {Bournaud}, {Fensch}, {Daddi},
  {Behrendt}, {Burkert}  \& {Richard}}{{Faure} et~al.}{2021}]{Faure21}
{Faure} B.,  {Bournaud} F.,  {Fensch} J.,  {Daddi} E.,  {Behrendt} M.,
  {Burkert} A.,   {Richard} J.,  2021, arXiv e-prints, \href
  {https://ui.adsabs.harvard.edu/abs/2021arXiv210111013F} {p. arXiv:2101.11013}

\bibitem[\protect\citeauthoryear{{Feldmann}, {Quataert}, {Hopkins},
  {Faucher-Gigu{\`e}re}  \& {Kere{\v{s}}}}{{Feldmann}
  et~al.}{2017}]{Feldmann17}
{Feldmann} R.,  {Quataert} E.,  {Hopkins} P.~F.,  {Faucher-Gigu{\`e}re} C.-A.,
   {Kere{\v{s}}} D.,  2017, \mn@doi [\mnras] {10.1093/mnras/stx1120}, \href
  {https://ui.adsabs.harvard.edu/abs/2017MNRAS.470.1050F} {470, 1050}

\bibitem[\protect\citeauthoryear{{Fensch} et~al.,}{{Fensch}
  et~al.}{2017}]{Fensch17}
{Fensch} J.,  et~al., 2017, \mn@doi [\mnras] {10.1093/mnras/stw2920}, \href
  {http://adsabs.harvard.edu/abs/2017MNRAS.465.1934F} {465, 1934}

\bibitem[\protect\citeauthoryear{{F{\"o}rster Schreiber} et~al.,}{{F{\"o}rster
  Schreiber} et~al.}{2011}]{Forster11}
{F{\"o}rster Schreiber} N.~M.,  et~al., 2011, \mn@doi [\apj]
  {10.1088/0004-637X/739/1/45}, \href
  {http://adsabs.harvard.edu/abs/2011ApJ...739...45F} {739, 45}

\bibitem[\protect\citeauthoryear{{F{\"o}rster Schreiber} et~al.,}{{F{\"o}rster
  Schreiber} et~al.}{2019}]{Forster19}
{F{\"o}rster Schreiber} N.~M.,  et~al., 2019, \mn@doi [\apj]
  {10.3847/1538-4357/ab0ca2}, \href
  {https://ui.adsabs.harvard.edu/abs/2019ApJ...875...21F} {875, 21}

\bibitem[\protect\citeauthoryear{Gabor \& Bournaud}{Gabor \&
  Bournaud}{2014}]{Gabor14}
Gabor J.~M.,  Bournaud F.,  2014, \mn@doi [Monthly Notices of the Royal
  Astronomical Society: Letters] {10.1093/mnrasl/slt139}, 437, L56–L60

\bibitem[\protect\citeauthoryear{{Genel} et~al.,}{{Genel}
  et~al.}{2012}]{Genel12}
{Genel} S.,  et~al., 2012, \mn@doi [\apj] {10.1088/0004-637X/745/1/11}, \href
  {http://cdsads.u-strasbg.fr/abs/2012ApJ...745...11G} {745, 11}

\bibitem[\protect\citeauthoryear{{Genel} et~al.,}{{Genel}
  et~al.}{2014}]{Genel14}
{Genel} S.,  et~al., 2014, \mn@doi [\mnras] {10.1093/mnras/stu1654}, \href
  {https://ui.adsabs.harvard.edu/abs/2014MNRAS.445..175G} {445, 175}

\bibitem[\protect\citeauthoryear{{Genzel} et~al.,}{{Genzel}
  et~al.}{2010}]{Genzel10}
{Genzel} R.,  et~al., 2010, \mn@doi [\mnras]
  {10.1111/j.1365-2966.2010.16969.x}, \href
  {http://cdsads.u-strasbg.fr/abs/2010MNRAS.407.2091G} {407, 2091}

\bibitem[\protect\citeauthoryear{{Genzel} et~al.,}{{Genzel}
  et~al.}{2011}]{Genzel11}
{Genzel} R.,  et~al., 2011, \mn@doi [\apj] {10.1088/0004-637X/733/2/101}, \href
  {http://adsabs.harvard.edu/abs/2011ApJ...733..101G} {733, 101}

\bibitem[\protect\citeauthoryear{Goldreich \& Lynden-Bell}{Goldreich \&
  Lynden-Bell}{1965}]{Goldreich65}
Goldreich P.,  Lynden-Bell D.,  1965, \mnras, 130, 28

\bibitem[\protect\citeauthoryear{{Guo}, {Giavalisco}, {Ferguson}, {Cassata}  \&
  {Koekemoer}}{{Guo} et~al.}{2012}]{Guo12}
{Guo} Y.,  {Giavalisco} M.,  {Ferguson} H.~C.,  {Cassata} P.,   {Koekemoer}
  A.~M.,  2012, \mn@doi [\apj] {10.1088/0004-637X/757/2/120}, \href
  {https://ui.adsabs.harvard.edu/abs/2012ApJ...757..120G} {757, 120}

\bibitem[\protect\citeauthoryear{{Guo} et~al.,}{{Guo} et~al.}{2015}]{Guo15}
{Guo} Y.,  et~al., 2015, \mn@doi [\apj] {10.1088/0004-637X/800/1/39}, \href
  {http://adsabs.harvard.edu/abs/2015ApJ...800...39G} {800, 39}

\bibitem[\protect\citeauthoryear{{Guo} et~al.,}{{Guo} et~al.}{2018}]{Guo18}
{Guo} Y.,  et~al., 2018, \mn@doi [\apj] {10.3847/1538-4357/aaa018}, \href
  {https://ui.adsabs.harvard.edu/abs/2018ApJ...853..108G} {853, 108}

\bibitem[\protect\citeauthoryear{Hopkins, Kereš  \& Murray}{Hopkins
  et~al.}{2013}]{Hopkins13}
Hopkins P.~F.,  Kereš D.,   Murray N.,  2013, \mn@doi [Monthly Notices of the
  Royal Astronomical Society] {10.1093/mnras/stt472}, 432, 2639–2646

\bibitem[\protect\citeauthoryear{{Hopkins}, {Kere{\v{s}}}, {O{\~n}orbe},
  {Faucher-Gigu{\`e}re}, {Quataert}, {Murray}  \& {Bullock}}{{Hopkins}
  et~al.}{2014}]{Hopkins14}
{Hopkins} P.~F.,  {Kere{\v{s}}} D.,  {O{\~n}orbe} J.,  {Faucher-Gigu{\`e}re}
  C.-A.,  {Quataert} E.,  {Murray} N.,   {Bullock} J.~S.,  2014, \mn@doi
  [\mnras] {10.1093/mnras/stu1738}, \href
  {https://ui.adsabs.harvard.edu/abs/2014MNRAS.445..581H} {445, 581}

\bibitem[\protect\citeauthoryear{Inoue, Dekel, Mandelker, Ceverino, Bournaud
  \& Primack}{Inoue et~al.}{2016}]{Inoue16}
Inoue S.,  Dekel A.,  Mandelker N.,  Ceverino D.,  Bournaud F.,   Primack J.,
  2016, \mn@doi [Monthly Notices of the Royal Astronomical Society]
  {10.1093/mnras/stv2793}, 456, 2052–2069

\bibitem[\protect\citeauthoryear{{Ivison}, {Richard}, {Biggs}, {Zwaan},
  {Falgarone}, {Arumugam}, {van der Werf}  \& {Rujopakarn}}{{Ivison}
  et~al.}{2020}]{Ivison20}
{Ivison} R.~J.,  {Richard} J.,  {Biggs} A.~D.,  {Zwaan} M.~A.,  {Falgarone} E.,
   {Arumugam} V.,  {van der Werf} P.~P.,   {Rujopakarn} W.,  2020, \mn@doi
  [\mnras] {10.1093/mnrasl/slaa046}, \href
  {https://ui.adsabs.harvard.edu/abs/2020MNRAS.495L...1I} {495, L1}

\bibitem[\protect\citeauthoryear{{Jog} \& {Solomon}}{{Jog} \&
  {Solomon}}{1984}]{Jog84}
{Jog} C.~J.,  {Solomon} P.~M.,  1984, \mn@doi [\apj] {10.1086/161597}, \href
  {http://adsabs.harvard.edu/abs/1984ApJ...276..114J} {276, 114}

\bibitem[\protect\citeauthoryear{Keller, Wadsley  \& Couchman}{Keller
  et~al.}{2015}]{Keller15}
Keller B.~W.,  Wadsley J.,   Couchman H. M.~P.,  2015, p.~11

\bibitem[\protect\citeauthoryear{{Kennicutt}}{{Kennicutt}}{1998}]{Kennicutt98}
{Kennicutt} Jr. R.~C.,  1998, \mn@doi [\apj] {10.1086/305588}, \href
  {http://cdsads.u-strasbg.fr/abs/1998ApJ...498..541K} {498, 541}

\bibitem[\protect\citeauthoryear{{Lagos} et~al.,}{{Lagos}
  et~al.}{2015}]{Lagos15}
{Lagos} C. d.~P.,  et~al., 2015, \mn@doi [\mnras] {10.1093/mnras/stv1488},
  \href {https://ui.adsabs.harvard.edu/abs/2015MNRAS.452.3815L} {452, 3815}

\bibitem[\protect\citeauthoryear{{Madau} \& {Dickinson}}{{Madau} \&
  {Dickinson}}{2014}]{Madau14}
{Madau} P.,  {Dickinson} M.,  2014, \mn@doi [\araa]
  {10.1146/annurev-astro-081811-125615}, \href
  {http://adsabs.harvard.edu/abs/2014ARA%26A..52..415M} {52, 415}

\bibitem[\protect\citeauthoryear{{Magdis} et~al.,}{{Magdis}
  et~al.}{2012}]{Magdis12}
{Magdis} G.~E.,  et~al., 2012, \mn@doi [\apj] {10.1088/0004-637X/760/1/6},
  \href {https://ui.adsabs.harvard.edu/abs/2012ApJ...760....6M} {760, 6}

\bibitem[\protect\citeauthoryear{Mandelker, Dekel, Ceverino, Tweed, Moody  \&
  Primack}{Mandelker et~al.}{2014}]{Mandelker14}
Mandelker N.,  Dekel A.,  Ceverino D.,  Tweed D.,  Moody C.~E.,   Primack J.,
  2014, \mn@doi [Monthly Notices of the Royal Astronomical Society]
  {10.1093/mnras/stu1340}, 443, 3675–3702

\bibitem[\protect\citeauthoryear{{Mandelker}, {Dekel}, {Ceverino}, {DeGraf},
  {Guo}  \& {Primack}}{{Mandelker} et~al.}{2017}]{Mandelker17}
{Mandelker} N.,  {Dekel} A.,  {Ceverino} D.,  {DeGraf} C.,  {Guo} Y.,
  {Primack} J.,  2017, \mn@doi [\mnras] {10.1093/mnras/stw2358}, \href
  {https://ui.adsabs.harvard.edu/abs/2017MNRAS.464..635M} {464, 635}

\bibitem[\protect\citeauthoryear{{Martizzi}, {Faucher-Gigu{\`e}re}  \&
  {Quataert}}{{Martizzi} et~al.}{2015}]{Martizzi15}
{Martizzi} D.,  {Faucher-Gigu{\`e}re} C.-A.,   {Quataert} E.,  2015, \mn@doi
  [\mnras] {10.1093/mnras/stv562}, \href
  {https://ui.adsabs.harvard.edu/abs/2015MNRAS.450..504M} {450, 504}

\bibitem[\protect\citeauthoryear{Mayer}{Mayer}{2016}]{Mayer16}
Mayer L.,  2016, The Astrophysical Journal Letters, p.~7

\bibitem[\protect\citeauthoryear{Moody, Guo, Mandelker, Ceverino, Mozena, Koo,
  Dekel  \& Primack}{Moody et~al.}{2014}]{Moody14}
Moody C.~E.,  Guo Y.,  Mandelker N.,  Ceverino D.,  Mozena M.,  Koo D.~C.,
  Dekel A.,   Primack J.,  2014, p.~11

\bibitem[\protect\citeauthoryear{{Muratov}, {Kere{\v{s}}},
  {Faucher-Gigu{\`e}re}, {Hopkins}, {Quataert}  \& {Murray}}{{Muratov}
  et~al.}{2015}]{Muratov15}
{Muratov} A.~L.,  {Kere{\v{s}}} D.,  {Faucher-Gigu{\`e}re} C.-A.,  {Hopkins}
  P.~F.,  {Quataert} E.,   {Murray} N.,  2015, \mn@doi [\mnras]
  {10.1093/mnras/stv2126}, \href
  {https://ui.adsabs.harvard.edu/abs/2015MNRAS.454.2691M} {454, 2691}

\bibitem[\protect\citeauthoryear{{Nelson} et~al.,}{{Nelson}
  et~al.}{2019}]{Nelson19}
{Nelson} D.,  et~al., 2019, \mn@doi [\mnras] {10.1093/mnras/stz2306}, \href
  {https://ui.adsabs.harvard.edu/abs/2019MNRAS.490.3234N} {490, 3234}

\bibitem[\protect\citeauthoryear{{Newman} et~al.,}{{Newman}
  et~al.}{2012}]{Newman12}
{Newman} S.~F.,  et~al., 2012, \mn@doi [\apj] {10.1088/0004-637X/761/1/43},
  \href {http://cdsads.u-strasbg.fr/abs/2012ApJ...761...43N} {761, 43}

\bibitem[\protect\citeauthoryear{{Noguchi}}{{Noguchi}}{1998}]{Noguchi98}
{Noguchi} M.,  1998, \mn@doi [\nat] {10.1038/32596}, \href
  {https://ui.adsabs.harvard.edu/abs/1998Natur.392..253N} {392, 253}

\bibitem[\protect\citeauthoryear{Noguchi}{Noguchi}{1999}]{Noguchi99}
Noguchi M.,  1999, \mn@doi [The Astrophysical Journal] {10.1086/306932}, 514,
  77–95

\bibitem[\protect\citeauthoryear{{Oklop{\v{c}}i{\'c}}, {Hopkins}, {Feldmann},
  {Kere{\v{s}}}, {Faucher-Gigu{\`e}re}  \& {Murray}}{{Oklop{\v{c}}i{\'c}}
  et~al.}{2017}]{Oklopcic17}
{Oklop{\v{c}}i{\'c}} A.,  {Hopkins} P.~F.,  {Feldmann} R.,  {Kere{\v{s}}} D.,
  {Faucher-Gigu{\`e}re} C.-A.,   {Murray} N.,  2017, \mn@doi [\mnras]
  {10.1093/mnras/stw2754}, \href
  {https://ui.adsabs.harvard.edu/abs/2017MNRAS.465..952O} {465, 952(O17)}

\bibitem[\protect\citeauthoryear{{Perez}, {Valenzuela}, {Tissera}  \&
  {Michel-Dansac}}{{Perez} et~al.}{2013}]{Perez13}
{Perez} J.,  {Valenzuela} O.,  {Tissera} P.~B.,   {Michel-Dansac} L.,  2013,
  \mn@doi [\mnras] {10.1093/mnras/stt1563}, \href
  {http://adsabs.harvard.edu/abs/2013MNRAS.436..259P} {436, 259}

\bibitem[\protect\citeauthoryear{{Perret}, {Renaud}, {Epinat}, {Amram},
  {Bournaud}, {Contini}, {Teyssier}  \& {Lambert}}{{Perret}
  et~al.}{2014}]{Perret14}
{Perret} V.,  {Renaud} F.,  {Epinat} B.,  {Amram} P.,  {Bournaud} F.,
  {Contini} T.,  {Teyssier} R.,   {Lambert} J.-C.,  2014, \mn@doi [\aap]
  {10.1051/0004-6361/201322395}, \href
  {http://adsabs.harvard.edu/abs/2014A%26A...562A...1P} {562, A1}

\bibitem[\protect\citeauthoryear{{Pillepich} et~al.,}{{Pillepich}
  et~al.}{2019}]{Pillepich19}
{Pillepich} A.,  et~al., 2019, \mn@doi [\mnras] {10.1093/mnras/stz2338}, \href
  {https://ui.adsabs.harvard.edu/abs/2019MNRAS.490.3196P} {490, 3196}

\bibitem[\protect\citeauthoryear{{Popping} et~al.,}{{Popping}
  et~al.}{2019}]{Popping19}
{Popping} G.,  et~al., 2019, \mn@doi [\apj] {10.3847/1538-4357/ab30f2}, \href
  {https://ui.adsabs.harvard.edu/abs/2019ApJ...882..137P} {882, 137}

\bibitem[\protect\citeauthoryear{{Rafikov}}{{Rafikov}}{2001}]{Rafikov01}
{Rafikov} R.~R.,  2001, \mn@doi [\mnras] {10.1046/j.1365-8711.2001.04201.x},
  \href {https://ui.adsabs.harvard.edu/abs/2001MNRAS.323..445R} {323, 445}

\bibitem[\protect\citeauthoryear{{Renaud} et~al.,}{{Renaud}
  et~al.}{2013}]{Renaud13}
{Renaud} F.,  et~al., 2013, \mn@doi [\mnras] {10.1093/mnras/stt1698}, \href
  {http://adsabs.harvard.edu/abs/2013MNRAS.436.1836R} {436, 1836}

\bibitem[\protect\citeauthoryear{{Renaud}, {Bournaud}  \& {Duc}}{{Renaud}
  et~al.}{2015}]{Renaud15}
{Renaud} F.,  {Bournaud} F.,   {Duc} P.-A.,  2015, \mn@doi [\mnras]
  {10.1093/mnras/stu2208}, \href
  {http://adsabs.harvard.edu/abs/2015MNRAS.446.2038R} {446, 2038}

\bibitem[\protect\citeauthoryear{{Romeo}, {Burkert}  \& {Agertz}}{{Romeo}
  et~al.}{2010}]{Romeo10}
{Romeo} A.~B.,  {Burkert} A.,   {Agertz} O.,  2010, \mn@doi [\mnras]
  {10.1111/j.1365-2966.2010.16975.x}, \href
  {https://ui.adsabs.harvard.edu/abs/2010MNRAS.407.1223R} {407, 1223}

\bibitem[\protect\citeauthoryear{{Rujopakarn} et~al.,}{{Rujopakarn}
  et~al.}{2019}]{Rujopakarn19}
{Rujopakarn} W.,  et~al., 2019, \mn@doi [\apj] {10.3847/1538-4357/ab3791},
  \href {https://ui.adsabs.harvard.edu/abs/2019ApJ...882..107R} {882, 107}

\bibitem[\protect\citeauthoryear{{Santini} et~al.,}{{Santini}
  et~al.}{2014}]{Santini14}
{Santini} P.,  et~al., 2014, \mn@doi [\aap] {10.1051/0004-6361/201322835},
  \href {https://ui.adsabs.harvard.edu/abs/2014A&A...562A..30S} {562, A30}

\bibitem[\protect\citeauthoryear{{Schmidt}}{{Schmidt}}{1959}]{Schmidt59}
{Schmidt} M.,  1959, \mn@doi [\apj] {10.1086/146614}, \href
  {http://cdsads.u-strasbg.fr/abs/1959ApJ...129..243S} {129, 243}

\bibitem[\protect\citeauthoryear{{Schroetter} et~al.,}{{Schroetter}
  et~al.}{2019}]{Schroetter19}
{Schroetter} I.,  et~al., 2019, \mn@doi [\mnras] {10.1093/mnras/stz2822}, \href
  {https://ui.adsabs.harvard.edu/abs/2019MNRAS.490.4368S} {490, 4368}

\bibitem[\protect\citeauthoryear{{Soko{\l}owska}, {Capelo}, {Fall}, {Mayer},
  {Shen}  \& {Bonoli}}{{Soko{\l}owska} et~al.}{2017}]{Sokolowska17}
{Soko{\l}owska} A.,  {Capelo} P.~R.,  {Fall} S.~M.,  {Mayer} L.,  {Shen} S.,
  {Bonoli} S.,  2017, \mn@doi [\apj] {10.3847/1538-4357/835/2/289}, \href
  {https://ui.adsabs.harvard.edu/abs/2017ApJ...835..289S} {835, 289}

\bibitem[\protect\citeauthoryear{{Soto} et~al.,}{{Soto} et~al.}{2017}]{Soto17}
{Soto} E.,  et~al., 2017, \mn@doi [\apj] {10.3847/1538-4357/aa5da3}, \href
  {https://ui.adsabs.harvard.edu/abs/2017ApJ...837....6S} {837, 6}

\bibitem[\protect\citeauthoryear{{Str{\"o}mgren}}{{Str{\"o}mgren}}{1939}]{Stromgren39}
{Str{\"o}mgren} B.,  1939, \mn@doi [\apj] {10.1086/144074}, \href
  {https://ui.adsabs.harvard.edu/abs/1939ApJ....89..526S} {89, 526}

\bibitem[\protect\citeauthoryear{{Tacconi} et~al.,}{{Tacconi}
  et~al.}{2010}]{Tacconi10}
{Tacconi} L.~J.,  et~al., 2010, \mn@doi [\nat] {10.1038/nature08773}, \href
  {https://ui.adsabs.harvard.edu/abs/2010Natur.463..781T} {463, 781}

\bibitem[\protect\citeauthoryear{{Tacconi} et~al.,}{{Tacconi}
  et~al.}{2018}]{Tacconi18}
{Tacconi} L.~J.,  et~al., 2018, \mn@doi [\apj] {10.3847/1538-4357/aaa4b4},
  \href {https://ui.adsabs.harvard.edu/abs/2018ApJ...853..179T} {853, 179}

\bibitem[\protect\citeauthoryear{{Tamburello}, {Mayer}, {Shen}  \&
  {Wadsley}}{{Tamburello} et~al.}{2015}]{Tamburello15}
{Tamburello} V.,  {Mayer} L.,  {Shen} S.,   {Wadsley} J.,  2015, \mn@doi
  [\mnras] {10.1093/mnras/stv1695}, \href
  {https://ui.adsabs.harvard.edu/abs/2015MNRAS.453.2490T} {453, 2490}

\bibitem[\protect\citeauthoryear{{Teyssier}}{{Teyssier}}{2002}]{Teyssier02}
{Teyssier} R.,  2002, \mn@doi [\aap] {10.1051/0004-6361:20011817}, \href
  {http://cdsads.u-strasbg.fr/abs/2002A%26A...385..337T} {385, 337}

\bibitem[\protect\citeauthoryear{{Teyssier}, {Chapon}  \&
  {Bournaud}}{{Teyssier} et~al.}{2010}]{Teyssier10}
{Teyssier} R.,  {Chapon} D.,   {Bournaud} F.,  2010, \mn@doi [\apjl]
  {10.1088/2041-8205/720/2/L149}, \href
  {http://cdsads.u-strasbg.fr/abs/2010ApJ...720L.149T} {720, L149}

\bibitem[\protect\citeauthoryear{{Toomre}}{{Toomre}}{1964}]{Toomre64}
{Toomre} A.,  1964, \mn@doi [\apj] {10.1086/147861}, \href
  {https://ui.adsabs.harvard.edu/abs/1964ApJ...139.1217T} {139, 1217}

\bibitem[\protect\citeauthoryear{{Truelove}, {Klein}, {McKee}, {Holliman},
  {Howell}  \& {Greenough}}{{Truelove} et~al.}{1997}]{Truelove97}
{Truelove} J.~K.,  {Klein} R.~I.,  {McKee} C.~F.,  {Holliman} II J.~H.,
  {Howell} L.~H.,   {Greenough} J.~A.,  1997, \mn@doi [\apjl] {10.1086/310975},
  \href {http://cdsads.u-strasbg.fr/abs/1997ApJ...489L.179T} {489, L179}

\bibitem[\protect\citeauthoryear{{Valentino} et~al.,}{{Valentino}
  et~al.}{2020}]{Valentino20}
{Valentino} F.,  et~al., 2020, \mn@doi [\apj] {10.3847/1538-4357/ab6603}, \href
  {https://ui.adsabs.harvard.edu/abs/2020ApJ...890...24V} {890, 24}

\bibitem[\protect\citeauthoryear{{Wuyts} et~al.,}{{Wuyts}
  et~al.}{2012}]{Wuyts12}
{Wuyts} S.,  et~al., 2012, \mn@doi [\apj] {10.1088/0004-637X/753/2/114}, \href
  {http://adsabs.harvard.edu/abs/2012ApJ...753..114W} {753, 114}

\bibitem[\protect\citeauthoryear{{Zanella} et~al.,}{{Zanella}
  et~al.}{2018}]{Zanella18}
{Zanella} A.,  et~al., 2018, \mn@doi [\mnras] {10.1093/mnras/sty2394}, \href
  {https://ui.adsabs.harvard.edu/abs/2018MNRAS.481.1976Z} {481, 1976}

\bibitem[\protect\citeauthoryear{{Zanella} et~al.,}{{Zanella}
  et~al.}{2019}]{Zanella19}
{Zanella} A.,  et~al., 2019, \mn@doi [\mnras] {10.1093/mnras/stz2099}, \href
  {https://ui.adsabs.harvard.edu/abs/2019MNRAS.489.2792Z} {489, 2792}

\bibitem[\protect\citeauthoryear{{Zolotov} et~al.,}{{Zolotov}
  et~al.}{2015}]{Zolotov15}
{Zolotov} A.,  et~al., 2015, \mn@doi [\mnras] {10.1093/mnras/stv740}, \href
  {https://ui.adsabs.harvard.edu/abs/2015MNRAS.450.2327Z} {450, 2327}

\makeatother
\end{thebibliography}


\appendix

\section{Stellar masses}
\label{App::stars}

\begin{figure*}
        \centering
         \includegraphics[width=8cm]{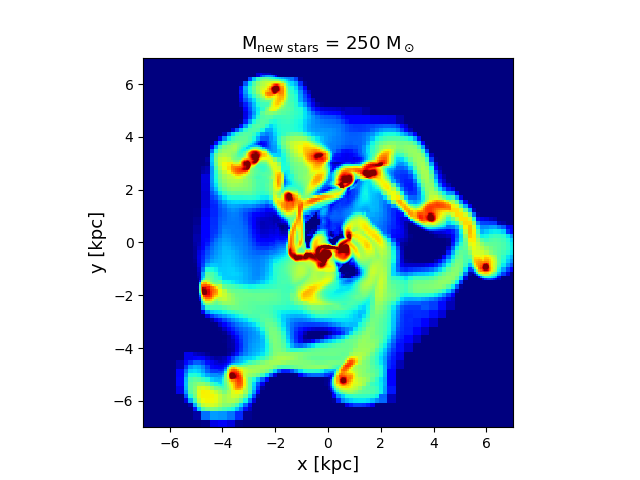}
         \includegraphics[width=8cm]{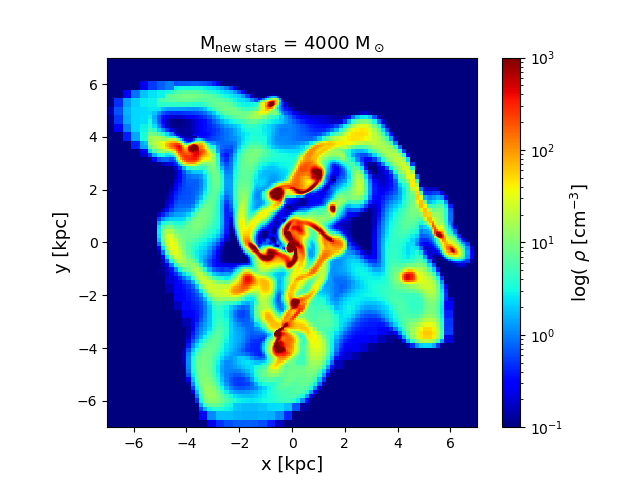} \\
          \includegraphics[width=8cm]{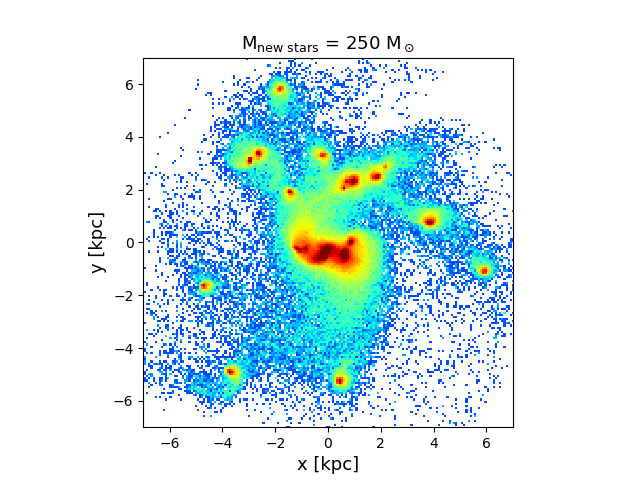}
         \includegraphics[width=8cm]{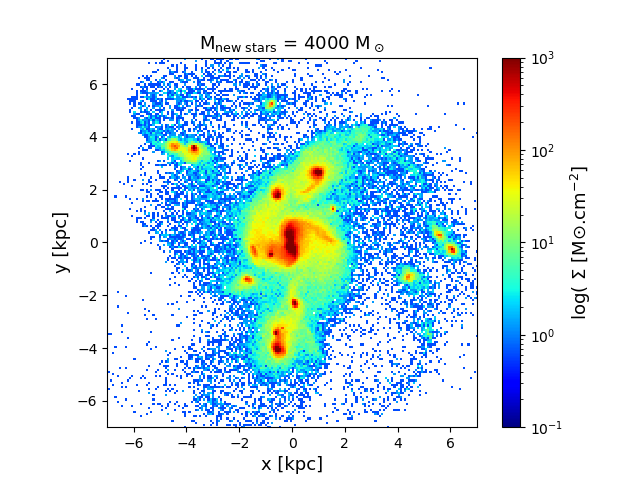} \\        
         \caption{{\bf Gas density maps and newly formed stars surface density maps of the F50 simulation with medium feedback after 175~Myr of evolution.}
             \label{fig::diff_star} }
        \end{figure*}

In Figure~\ref{fig::diff_star} we show the gas density maps and newly formed stars surface density maps of the F50 simulation with medium feedback with two different mass for the newly formed stars, namely 250~M$_\odot$ and 4000~M$_\odot$, after 175~Myr of evolution. Note that the disc instability being driven by stochasticity, there can not be a one-to-one match in the maps. The ratio between the two stellar particles masses being 16, for better comparison the stellar density map of the F50 run with 250~M$_\odot$ mass stellar particles (bottom left map) uses 1 stellar particle out of 16, which we assigned a 16 times higher mass. We do not see a major difference in the type of fragmentation in the disks. 

\section{Toomre maps and profiles}
\label{App::toomre}

\begin{figure*}
        \centering
         \includegraphics[width=8cm]{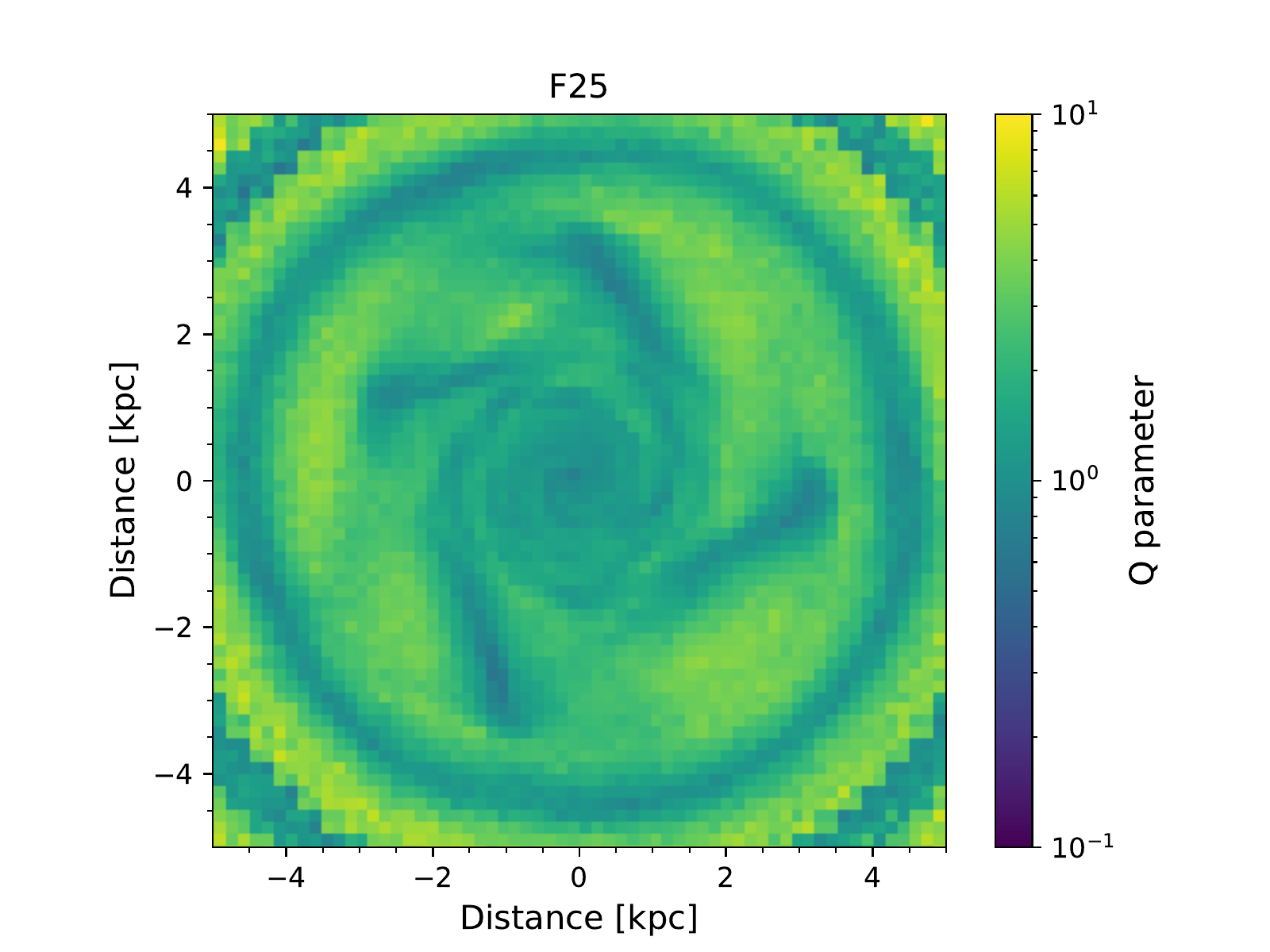}
         \includegraphics[width=8cm]{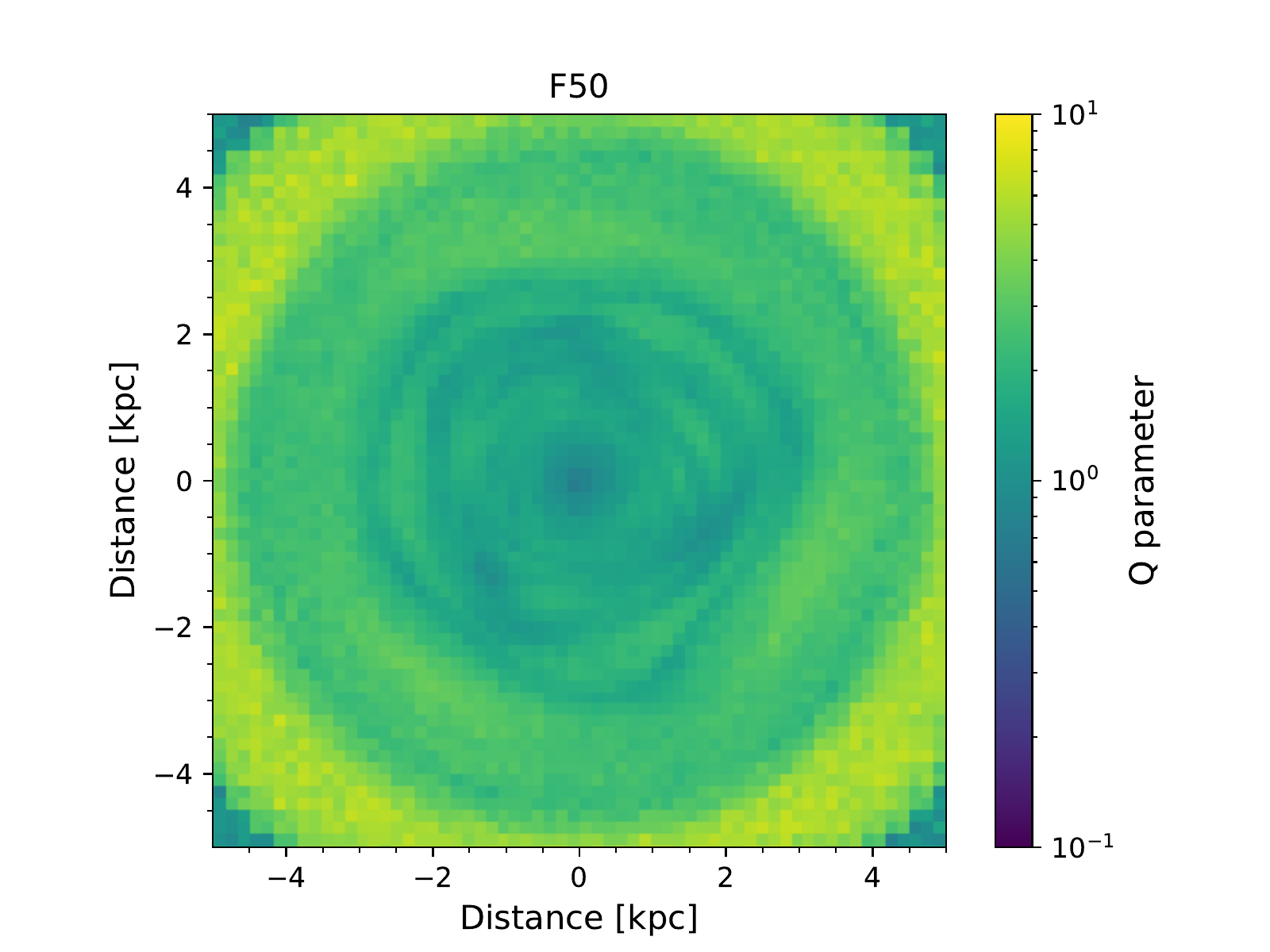}\\
         \includegraphics[width=9cm]{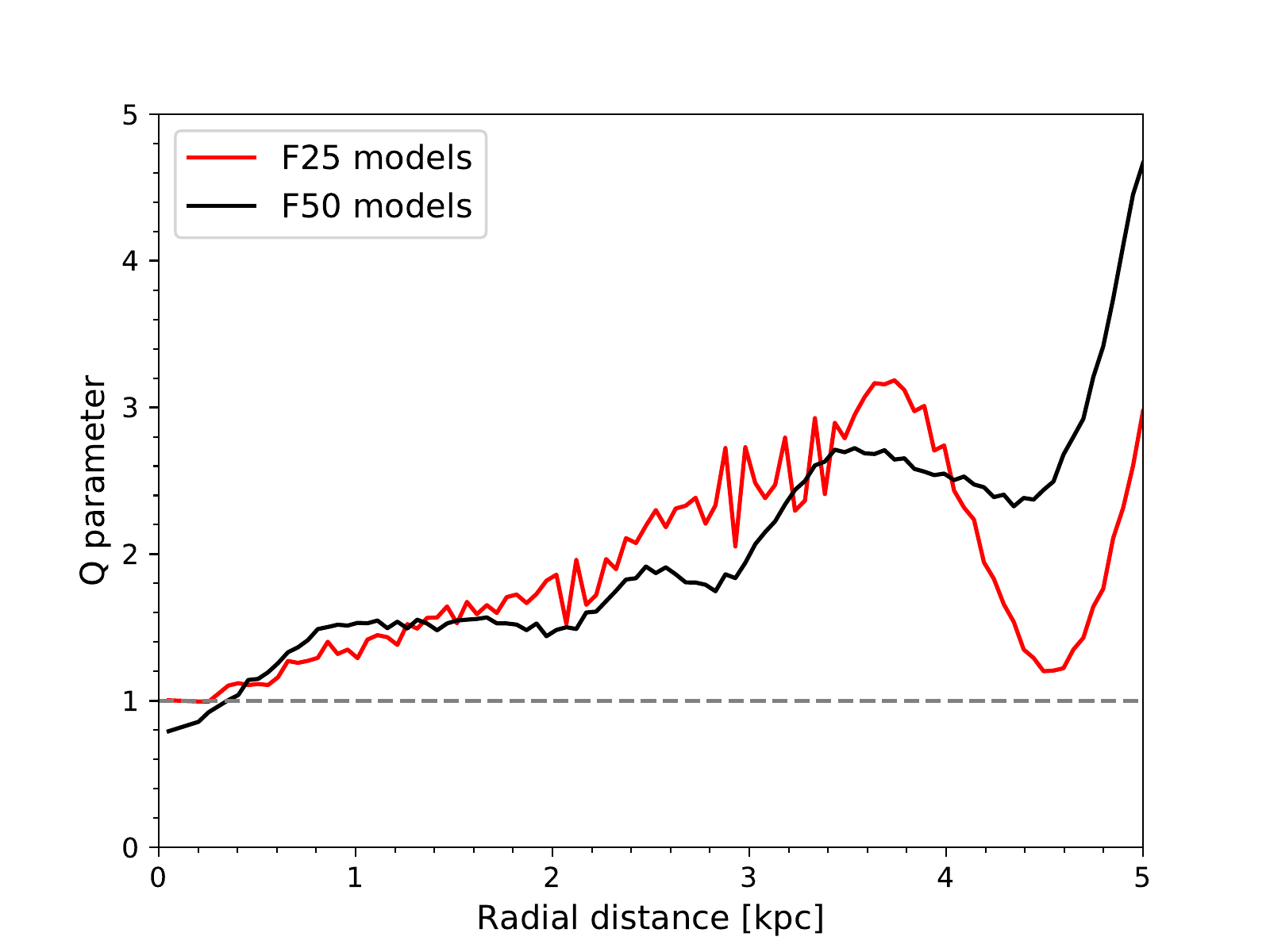}
         \caption{Maps and radial profile of the Toomre $Q$ parameter at the end of the relaxation phase.
             \label{fig:A_Toomre} }
        \end{figure*}

\begin{figure*}
        \centering
         \includegraphics[width=8cm]{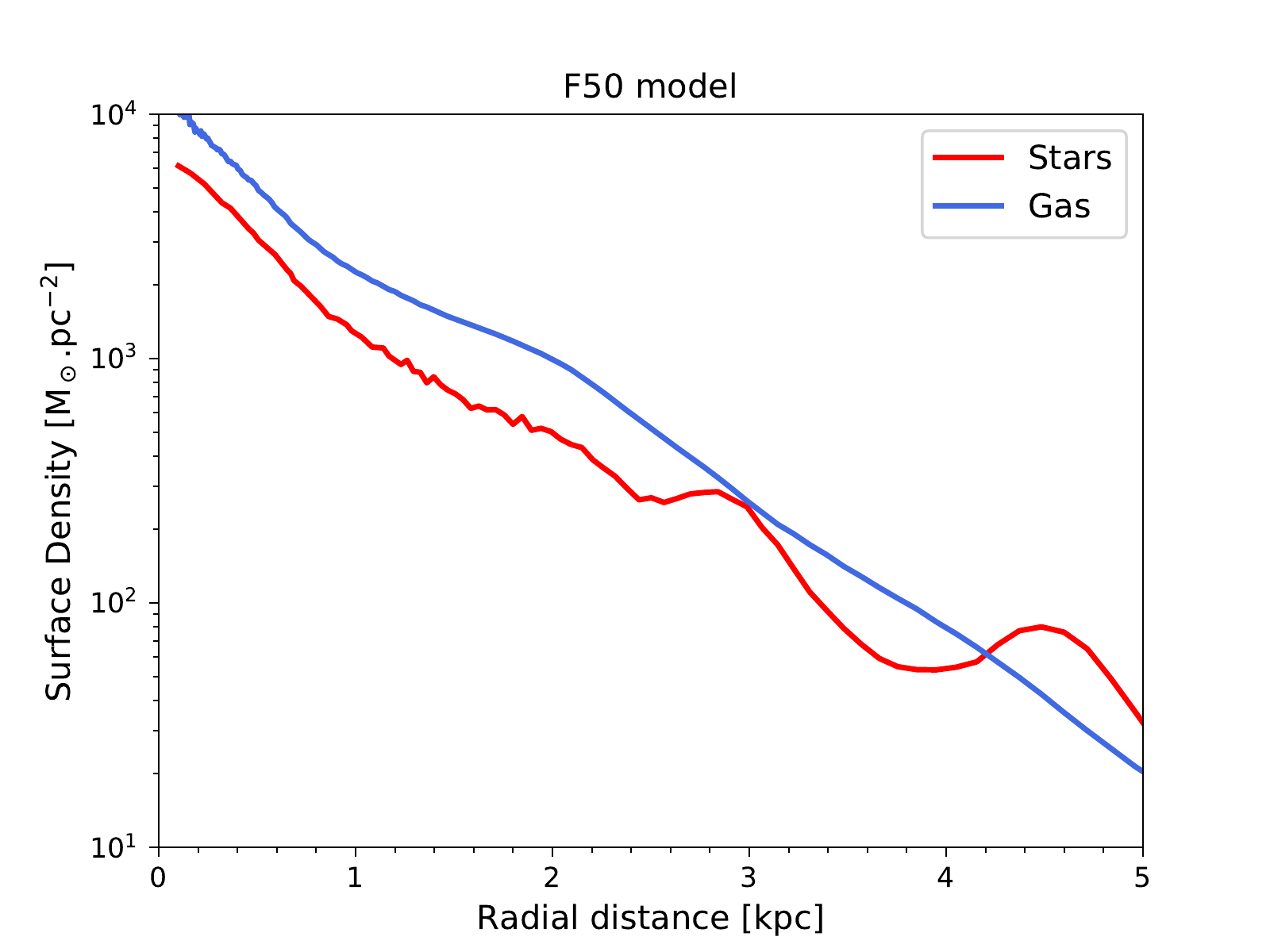}
         \caption{Maps and radial profile of the Toomre $Q$ parameter at the end of the relaxation phase.
             \label{fig::A_surfaced} }
        \end{figure*}

In Figure~\ref{fig:A_Toomre} we show maps of the Toomre $Q$ parameter and its radial profile for the F25 and F50 runs, obtained after the relaxation phase. The parameter $Q$ is defined by:

\begin{align}
    Q = \frac{\kappa \sqrt{c_s^2 + \sigma_\mathrm{gas}^2}}{\pi G (\Sigma_\mathrm{gas} + \Sigma_\star)}
\end{align}

with $c_s$ the sound speed, $\sigma$ and $\Sigma$ the velocity dispersion and the surface density of the gas or the stars measured at the 160~pc scale, and $\kappa$ the epicyclic frequency, defined by $\kappa(r)^2 = \frac{2\Omega}{r} \frac{\mathrm{d}}{\mathrm{d}r}(r^2 \Omega)$. In Figure~\ref{fig::A_surfaced} we show gas and stellar surface density profile of the F50 runs at the end of the relaxation phase, which can be approximated by exponential profiles.

\bsp	
\label{lastpage}
\end{document}